%% file: paper.tex
\documentclass[12pt]{article}

\pdfoutput=1
\usepackage{color}
\usepackage{epsfig, palatino}
\usepackage{pstricks,pst-node,pst-tree}
\usepackage{epic}
\usepackage{mathrsfs}
\usepackage{ae} 
\usepackage[T1]{fontenc}
\usepackage[ansinew]{inputenc}
\usepackage{amsmath}
\usepackage{amssymb}
\usepackage{graphicx}
\usepackage{ulem}
\usepackage{color}
\definecolor{darkblue}{cmyk}{0.9,0.9,0,0}
\usepackage[colorlinks=true,linkcolor=darkblue,citecolor=darkblue,urlcolor=darkblue]{hyperref}
\usepackage{cite}
\usepackage{hyperref}
\usepackage{wasysym}
\usepackage{varioref}
\usepackage{makeidx}
\usepackage[english]{babel}
\usepackage{simplewick}
\usepackage{array}
\usepackage[font={small}]{caption}

\newcommand{\comment}[1]{}

\newcommand{\beq}{\begin{equation}}
\newcommand{\eeq}{\end{equation}}
\newcommand{\beqq}{\begin{equation*}}
\newcommand{\eeqq}{\end{equation*}}
\newcommand\beqa{\begin{eqnarray}}
\newcommand\eeqa{\end{eqnarray}}
\newcommand\beqaa{\begin{eqnarray*}}
\newcommand\eeqaa{\end{eqnarray*}}
\newcommand\bea{\begin{array}}
\newcommand\eea{\end{array}}

\newcommand{\nn}{\nonumber}

\newcommand{\neqa}{\nonumber\end{eqnarray}} 
\newcommand{\la}[1]{\label{#1}}

\renewcommand{\d}{\partial}

\newcommand{\<}{{\langle}}
\renewcommand{\>}{{\rangle}}

\newcommand{\re}{\relax{\rm I\kern-.18em R}}

\renewcommand{\sp}{p\hspace{-.40em}/}

\newcommand{\Blue}[1]{{\color{blue}#1\color{black}}}
\newcommand{\Red}[1]{{\color{red}#1\color{black}}}

\newenvironment{Ceq}
 {
  }

\def\XXint#1#2#3{{\setbox0=\hbox{$#1{#2#3}{\int}$}
\vcenter{\hbox{$#2#3$}}\kern-.5\wd0}}

\def\su2{{SU(2)}}

\def\[{\left[}
\def\]{\right]}

\def\({\left(}
\def\){\right)}
\def\[{\left[}
\def\]{\right]}

\def\<{\langle}
\def\>{\rangle}

\def\i2{\frac{i}{2}}

\def\spi{\relax{\rm \pi\kern-0.5em /}}
\def\sA{\relax{\rm A\kern-0.5em /}}
\def\sp{\relax{\rm p\kern-0.5em /}}
\def\sd{\relax{\rm \d\kern-0.5em /}}
\def\sk{\relax{\rm k\kern-0.5em /}}
\def\sn{\relax{\rm n\kern-0.5em /}}
\def\sl{\relax{\rm l\kern-0.5em /}}
\def\sP{\relax{\rm P\kern-0.7em /}}
\def\sBethe{\relax{\rm \Bethe\kern-0.5em /}}

\def\One{1\hskip-.16cm1}

\def\cR{{\cal R}}

\def\cP{{\cal P}}

\def\cW{{\cal W}}

\def\cQ{{\cal Q}}
\def\cD{{\cal D}}
\def\2F1{\,_2{\rm F}_1}

\makeindex

\begin{document}

\thispagestyle{empty}

\renewcommand{\thefootnote}{\fnsymbol{footnote}}
\setcounter{page}{1}
\setcounter{footnote}{0}
\setcounter{figure}{0}
\begin{center}
$$$$
{\Large\textbf{\mathversion{bold}
OPE for all Helicity Amplitudes
}\par}

\vspace{1.0cm}

\textrm{Benjamin Basso$^\text{\tiny 1}$, Jo\~ao Caetano$^\text{\tiny 2,3,4,5}$, Luc\'ia C\'ordova$^\text{\tiny 2,3}$, Amit Sever$^\text{\tiny 6}$ and Pedro Vieira$^\text{\tiny 2}$}
\\ \vspace{1.2cm}
\footnotesize{\textit{
$^\text{\tiny 1}$Laboratoire de Physique Th\'eorique, \'Ecole Normale Sup\'erieure, Paris 75005, France\\
$^\text{\tiny 2}$Perimeter Institute for Theoretical Physics,
Waterloo, Ontario N2L 2Y5, Canada\\
$^\text{\tiny 3}$Department of Physics and Astronomy \& Guelph-Waterloo Physics Institute, University of Waterloo, Waterloo, Ontario N2L 3G1, Canada\\
$^\text{\tiny 4}$Mathematics Department, King's College London, The Strand, London WC2R 2LS, UK\\
$^\text{\tiny 5}$Centro de F$\acute{\imath}$sica do Porto, Departamento de F$\acute{\imath}$sica e Astronomia,
Faculdade de Ci$\hat{e}$ncias da Universidade do Porto, Rua do Campo Alegre 687, 4169-007 Porto, Portugal\\
$^\text{\tiny 6}$School of Physics and Astronomy, Tel Aviv University, Ramat Aviv 69978, Israel
}  
\vspace{4mm}
}

\par\vspace{1.5cm}

\textbf{Abstract}\vspace{2mm}
\end{center}

We extend the Operator Product Expansion (OPE) for scattering amplitudes in planar ${\cal N}=4$ SYM to account for all possible helicities of the external states. This is done by constructing a simple map between helicity configurations and so-called {\it charged pentagon transitions}. These OPE building blocks are generalizations of the bosonic pentagons entering MHV amplitudes and they can be
bootstrapped at finite coupling from the integrable dynamics of the color flux tube. A byproduct of our map is a simple realization of parity in the super Wilson loop picture. 
 
\noindent

\setcounter{page}{1}
\renewcommand{\thefootnote}{\arabic{footnote}}
\setcounter{footnote}{0}

 \def\nref#1{{(\ref{#1})}}

\newpage

\tableofcontents

\parskip 5pt plus 1pt   \jot = 1.5ex

\section{Introduction}

\begin{figure}[t]
\centering
\def\svgwidth{15cm}
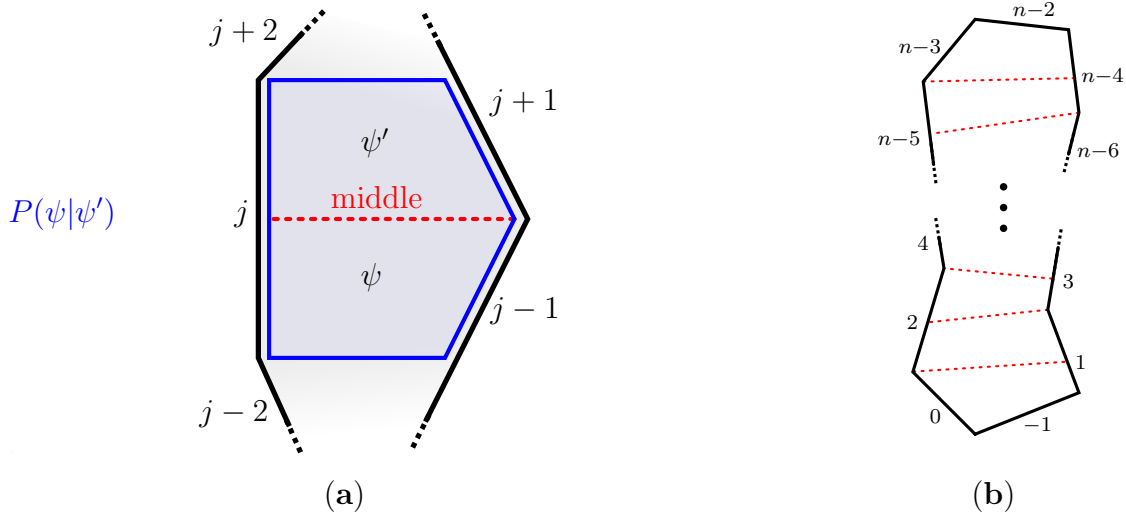
\caption{{\bf a}) The pentagon transitions are the building blocks of null polygonal Wilson loops. They represent the transition $\psi\rightarrow\psi'$ undergone by the flux-tube state as we move from one square to the next in the OPE decomposition. This breaking into squares is univocally defined by specifying the middle (or inner dashed) edge of the pentagon to be $Z_\text{middle}\propto\<j-2,j,j+2,j-1\>Z_{j+1}-\<j-2,j,j+2,j+1\>Z_{j-1}$. {\bf b})
In the OPE-friendly labelling of edges, adopted in this paper, the middle edge of the $j$-th pentagon ends on the $j$-th edge. As a result, the very bottom edge is edge $-1$ while the very top one is edge $n-2$. The map between the OPE index $j$ and the more common cyclic index $j_\text{cyc}$ reads $j_\text{cyc}=\frac{3}{4}-\frac{1}{4} (-1)^j (2 j+3)$ mod $n$.
}\label{pentagontransition}
\end{figure}

Within the so-called \textit{pentagon approach} for null polygonal Wilson loops in conformal gauge theories, one breaks up a null polygon into much simpler building blocks called \textit{pentagon transitions} $P(\psi|\psi')$.
These ones govern the transitions between two eigenstates of the color flux tube -- see figure \ref{pentagontransition}.\textbf{a} -- and provide a representation of the Wilson loop $\mathcal{W}_n$ -- or more precisely of the finite ratio of loops \cite{short} depicted in figure \ref{calW} -- in the form of an infinite sum over all OPE channels,
\beq\la{decomposition}
\mathcal{W}_n=\sum_{\psi_i}  P(0|\psi_1) P({\psi}_1|\psi_2)\dots P({\psi}_{n-6}|\psi_{n-5})P({\psi}_{n-5}|0) \,e^{\sum_{j}\(-E_j\tau_j+ip_j\sigma_j+im_j\phi_j\)}\, ,
\eeq
with $\{\tau_i,\sigma_i,\phi_i\}$ a base of conformal cross ratios, which receive individually meaning of time, space and angle in the $i$'th OPE channel 
\cite{short,moreOPE}. 

What makes this decomposition extremely powerful in planar  $\mathcal{N}=4$ SYM theory is that all of its building blocks can be computed at any value of the coupling thanks to the integrability of the underlying theory. Namely, the flux tube spectrum is under total control \cite{BenDispPaper} and the pentagon transitions can be bootstrapped  \cite{short,data,2pt,GluonPaper,ToAppear} following (a slightly modified version of) the standard form factor program for integrable theories. 

Through the celebrated duality between null polygonal Wilson loops and scattering amplitudes \cite{AM,AmplitudeWilson}, the decomposition (\ref{decomposition}) also provides a fully non-perturbative representation of the so-called Maximal Helicity Violating (MHV) gluon scattering amplitudes in planar $\mathcal{N}=4$ SYM theory. 

In this paper we will argue that a suitable generalization of the pentagon transitions into super or \textit{charged} pentagon transitions allows one to describe all amplitudes, for any number of external particles with arbitrary helicities and at any value of the 't Hooft coupling. 

While the key ingredient in having an OPE expansion such as (\ref{decomposition}) is conformal symmetry, a central ingredient in the charged pentagon approach will be supersymmetry. 

The idea of charging the pentagons is not entirely new, and already appeared in \cite{data} where certain charged transitions were introduced and successfully compared against N$^k$MHV amplitudes. More recently, further charged transitions were bootstrapped and matched with amplitudes in \cite{andrei1,GluonPaper,andrei2}. 

The aim of this paper is to complete this picture by proposing a simple map between all possible helicity amplitudes and all the ways charged pentagons can be patched together into an OPE series like (\ref{decomposition}). An interesting outcome of this charged pentagons analysis is a simple proposal for how parity acts at the level of the super Wilson loop, which, as far as we are aware, was not known before. 

 \section{The Charged Pentagon Program} 
In the dual Wilson loop picture, N$^k$MHV amplitudes are computed by a super Wilson loop decorated by adjoint fields inserted on the edges and cusps~\cite{superloopskinner,superloopsimon}. It is this super loop that we want to describe within the pentagon approach.

At first, let us first ask ourselves what would be a natural extension of (\ref{decomposition}) that allows for some regions of the loop to be \textit{charged} due to the insertion of these extra fields. The minimal modification one could envisage is to generalize the pentagon transitions to super pentagon transitions or \textit{charged} transitions, in which $P(\psi|\psi')$ stands as the bottom component. As for the $\mathcal{N}=4$ on-shell super field, a pentagon would naturally come in multiple of five components
\beq
\mathbb{P}={P}+\chi^A{P}_A+\chi^A\chi^B{P}_{AB}+\,\chi^A\chi^B\chi^C{P}_{ABC}+\chi^A\chi^B\chi^C\chi^D{P}_{ABCD} \,,\la{superP}
\eeq
where $\chi$ is a Grassmann parameter, $A=1,2,3,4$ an $R$-charge index, and where, for sake of clarity, we have suppressed the states $\psi$ and $\psi'$. With these charged transitions at hand, we could now imagine building up charged polygons such as 
\beqa
\cP_A\circ\cP^A&\equiv& \sum_{\psi_1} P_A(0|\psi_1) P^A(\psi_1|0)  e^{-E_1 \tau_1+\dots}\, ,\nn\\
\cP_{AB}\circ\cP\circ\cP^{AB}&\equiv&\!\! \sum_{\psi_1,\psi_2} \!\!P_{AB}(0|\psi_1) P(\psi_1|\psi_2) P^{AB}(\psi_2|0) e^{-E_1 \tau_1+\dots }\, ,\la{examples} \\
\cP_{AB}\circ\cP_{CD}\circ\cP^{AB}\circ\cP^{CD}&\equiv&\!\! \sum_{\psi_1,\psi_2,\psi_3} \!\!\!P_{AB}(0|\psi_1) P_{CD}(\psi_1|\psi_2) P^{AB}(\psi_2|\psi_3)P^{CD}(\psi_3|0) e^{-E_1 \tau_1+\dots } \nn
\eeqa
and so on. Here, an upper index represents a contraction with an epsilon tensor. Namely, we use $P^A=\epsilon^{ABCD}P_{BCD}$, $P^{AB}=\epsilon^{ABCD}P_{CD}$ and $P^{ABC}=\epsilon^{ABCD}P_{D}$ to compress the expressions above.

The most obvious change with respect to the MHV case is that $R$-charge conservation now forbids some of the processes which were previously allowed and vice-versa. For instance, in the creation amplitude $P_{AB}(0|...)$ we can produce a scalar $\phi_{AB}$ out of the vacuum, since this excitation has quantum numbers that match those of the charged pentagon. At the same time, neutral states such as the vacuum or purely gluonic states -- which appeared in the non-charged transitions -- can no longer be produced by this charged pentagon.

What stays the same is that all these charged transitions can be bootstrapped using integrability -- as much as their bosonic counterparts. The \textit{scalar} charged transition $P_{AB}$ and the \textit{gluon} charged transition $P_{ABCD}$, for instance, already received analysis of this sort in \cite{data,GluonPaper}.\footnote{Both were denoted by $P_*$ in these works.} The fermonic charged transitions, $P_A$ and $P_{ABC}$, were more recently constructed in \cite{andrei1,andrei2}.

The super pentagon hypothesis~(\ref{superP}) and its OPE corollary~(\ref{examples}) are the two main inputs in the charged pentagon program for helicity amplitudes. In the rest of this section we present a simple counting argument supporting the equivalence between super OPE series and super amplitudes.

The important point is that not all the N$^k$MHV amplitudes are independent. Because of supersymmetry, many of them get linked together by means of so-called SUSY Ward identities. At given number $n$ of particles, there is a basis of $\mathcal{N}(k,n)$ amplitudes in terms of which one can linearly express all the remaining ones.

The problem of eliminating this redundancy, such as to count the $\mathcal{N}(k,n)$ independent amplitudes, was beautifully analyzed in \cite{Elvang}. As explained below, the very same counting applies to inequivalent super OPE series like~(\ref{examples}).

Counting the number of super OPE series is relatively easy:

At first, one notices that the $R$-charge of a polygon is always a multiple of four, as a consequence of $SU(4)$ symmetry. The first two cases in (\ref{examples}), for instance, involve charged pentagons with a total of $4$ units, as for NMHV amplitudes, while the last example in~(\ref{examples}) has a total of $8$ units of charge, and should thus be related to N$^2$MHV amplitudes.

In the NMHV case, the amount of charge in each of the $n-4$ pentagons uniquely specifies the super OPE series and there is clearly $(n-1)(n-2)(n-3)(n-4)/4!$ ways of distributing four units of charge between the $n-4$ pentagons in our tessellation. Precisely this number is reported for $\mathcal{N}(1,n)$ in \cite{Elvang}, see discussion below (3.12) therein.

This kind of partitions no longer enumerate all cases starting with N$^2$MHV amplitudes. For instance, there are three independent ways of charging all the four pentagons of an octagon with two units of charge,

\beq
\cP_{AB}\circ\cP_{CD}\circ\cP^{AB}\circ\cP^{CD}\,,\quad\cP_{AB}\circ\cP^{AB}\circ\cP_{CD}\circ\cP^{CD}\,,\quad\cP_{AB}\circ\cP_{CD}\circ\cP^{CD}\circ\cP^{AB}\,,
\eeq
with the last line in (\ref{examples}) being one of them. 
(We can understand this as coming from the three possible irreducible representations in ${\bf 6} \otimes {\bf 6}$ or, equivalently, as the three inequivalent ways of forming singlets in ${\bf 6} \otimes {\bf 6} \otimes {\bf 6}\otimes {\bf 6}$.) Therefore, to count the number of N$^2$MHV charged polygons we have to consider not only the number of ways of distributing eight units of charge within four pentagons but also to weight that counting by the number of inequivalent contractions of all the R-charge indices. Remarkably, this counting is identical to the one found in~\cite{Elvang} based on analysis of the SUSY Ward identities. This is particularly obvious when looking at Table 1 in \cite{Elvang} where the number of independent N$^2$MHV components for $8$ and $9$ particles is considered.
\footnote{The weight $3=\mathcal{S}_{\lambda=[2,2,2,2]}$ in their table is precisely the one explained in our above discussion.}
In sum, our construction in (\ref{examples}) generates precisely $\mathcal{N}(2,8)=105,\,\mathcal{N}(2,9)=490,\dots$ different N$^2$MHV objects, in perfect agreement with the number of independent components arising from the study of the SUSY Ward identities.

It is quite amusing that the notation in \cite{Elvang} with a partition vector $\lambda=[\lambda_1,\dots,\lambda_{n-4}]$ seems perfectly tailored to describe the charged pentagon approach where we have $n-4$ pentagons with charges $\lambda_i \in \{0,1,2,3,4\}$.  It also guarantees that the most general N$^k$MHV counting works the same for both amplitudes and OPE series, and concludes this analysis. The next step is to endow the charged pentagon construction with a precise dictionary between charged polygons and helicity configurations of scattering amplitudes.

\section{The Map}

A compact way of packaging together all helicity amplitudes is through a generating function, also known as the super Wilson loop \cite{superloopskinner,superloopsimon} 
\beq
W_\text{super}=W_\text{MHV}+ \eta_i^1 \eta_j^2 \eta_k^3 \eta_l^4\,\, W_\text{NMHV}^{(ijkl)} + \eta_i^1 \eta_j^2 \eta_k^3 \eta_l^4  \eta_{m}^1 \eta_{n}^2 \eta_{o}^3 \eta_{p}^4\,\,W_\text{N$^2$MHV}^{(ijkl)(mnop)}  + \dots   \la{superdefinition}
\eeq
where $W_\text{N$^k$MHV}$ is the N$^k$MHV amplitude divided by the Parke-Taylor MHV factor. Here, the $\eta$'s are the dual Grassmann variables \cite{etas,Dualsuperconforma}. They transform in the fundamental of the $SU(4)$ R-symmetry, as indicated by their upper index $A=1,2,3,4$, and are associated to the edges of the polygon, indicated by the lower index $i=-1,0,1,\dots,n-2$. 

Throughout this paper we shall be using a rather unorthodox labelling of the edges of the polygon, which is represented in figure~\ref{pentagontransition}. Namely, we number the edges from bottom to top, with even numbers on one side and odd numbers on the other, like door numbers within a street. Given that we think of the Wilson loop as a sequence of flux-tube states propagating down this \textit{street},  this is the most natural labelling from the OPE viewpoint. 
It makes it particularly simple to locate the $j$-th pentagon in the tessellation: it is the pentagon whose middle edge ends on edge~$j$. 
The map between this labelling and the conventional cyclic ordering is explained in the caption of figure~\ref{pentagontransition}.
\begin{figure}[t]
\centering
\def\svgwidth{12cm}
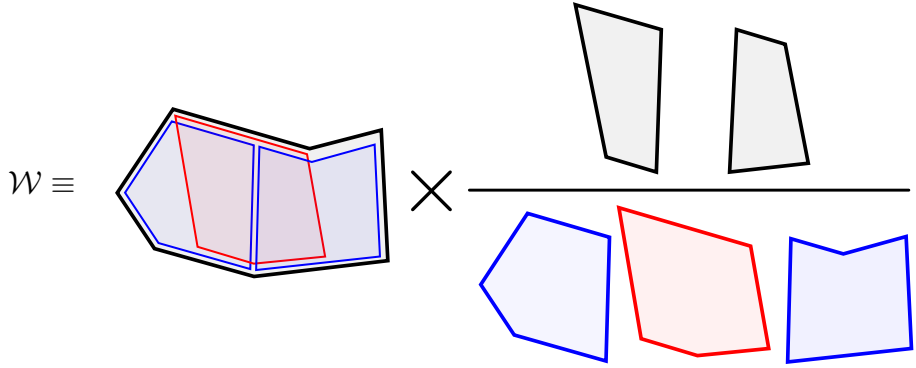
\caption{We study the conformally invariant and finite ratio $\mathcal W$ introduced in \cite{short}. It is obtained by dividing the expectation value of the super Wilson loop by all the pentagons in the decomposition and by multiplying it by all the middle squares. The twistors that define these smaller pentagons are either the twistors of the original polygon (an heptagon in this figure) or the middle twistors described in figure \ref{pentagontransition}.a (in the above figure there are three distinct middle twistors, for instance).}\label{calW}
\end{figure}

The super loop (\ref{superdefinition}) has UV suppression factors associated to its cusps.\footnote{These UV divergences are T-dual to the IR divergences of the on-shell amplitudes.} One can find in the literature several different ways of renormalizing the loop, such as to remove these factors.
The one most commonly used is the {\it ratio function} $\cR\equiv W_\text{super}/W_\text{MHV}$, first introduced in \cite{Dualsuperconforma}. For our discussion, however,  the OPE renormalization is better suited: it is obtained by dividing the super loop (\ref{superdefinition}) by all the pentagons in its decomposition and by multiplying it by all the middle squares \cite{data}
\beq
\cW\equiv W_\text{super}/{\rm w}\qquad\text{with}\qquad {\rm w} \equiv {\(\prod_{i=1}^{n-4}\< W_{i\text{'th pentagon}}\>\)}/{\(\prod_{i=1}^{n-5}\< W_{i\text{'th middle square}}\>\)}\, ,
\la{cWdef}
\eeq
as shown in figure \ref{calW}. The ratio function $\cR$ and the loop $\cW$ are then easily found to be related to each other by $\cR= \cW/\cW_\text{MHV}$. They are essentially equivalent, being both finite and conformally invariant functions of the $\eta$'s and shape of the loop, but only $\cR$ is cyclic invariant. 

\subsection{The Direct Map}

Our goal in this section is to find the map between the different ways of gluing the charged transitions together, as in (\ref{examples}), and the components of the super loop (\ref{cWdef}). Put differently, we would like to find a map between the $\eta$'s and the $\chi$'s such that $\cW$ in~(\ref{cWdef}) also admits the expansion
\beq
\cW =\cP\circ\cP\circ\dots\circ\cP+\chi_1^1\chi_1^2\chi_1^3\chi_1^4\,\cP_{1234}\circ\cP\circ\dots\circ\cP+\chi_1^1\chi_1^2\chi_1^3\chi_2^4\,\cP_{123}\circ\cP_4\circ\dots\circ\cP+\dots\la{superloopinPs}
\eeq
in terms of the $\chi$'s.

There are two important properties of the super loop that will be relevant to our discussion.

First, recall that an $\eta$ is associated to an edge of the polygon while a $\chi$ is associated to a pentagon. As such, there are many more terms in the $\eta$-expansion~(\ref{superdefinition}) or (\ref{cWdef}) of the super loop than there are in the $\chi$-expansion~(\ref{superloopinPs}). 
This is no contradiction, however. The reason is that the $\eta$-components are not all linearly independent, since, as mentioned before, they are subject to SUSY Ward identities. On the contrary, the $\chi$-components \textit{all} have different OPE interpretation and, in line with our previous discussion, should be viewed as defining a basis of independent components for the amplitudes. In other words, the map between $\chi$- and $\eta$-components is not bijective if not modded out by the SUSY Ward identities. We can then think of the $\chi$-decomposition as a natural way of getting rid of SUSY redundancy. 

Second, the $\eta$-components of $\cW$ are not `pure numbers', since they carry weights under the little group; e.g., upon rescaling of the twistor $Z_1\to\alpha\, Z_1$ the component $\cW^{1123}_\text{NMHV}$ transforms as $\cW^{1123}_\text{NMHV}\to\cW^{1123}_\text{NMHV}/\alpha^2$. These helicity weights cancel against those of the $\eta$'s, so that $\cW$ is weight free in the end. In contrast, the components in~(\ref{superloopinPs}), as well as the corresponding $\chi$'s, are taken to be weightless. With this choice, the $\chi$-components coincide with the ones predicted from integrability with no additional weight factors. 

\begin{figure}[t]
\centering
\def\svgwidth{12cm}
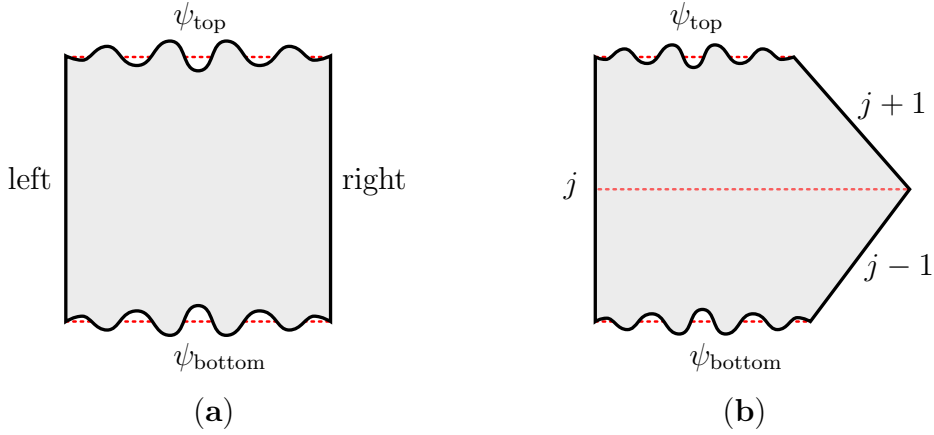
\caption{{\bf a}) Any square in the OPE decomposition stands for a transition from the state at its bottom ($\psi_\text{bottom}$) to the state at its top ($\psi_\text{top}$). This transition is generated by a conformal symmetry of the right and left edges of that square (conjugate to the flux time $\tau$). {\bf b}) Similarly, the super pentagon $\mathbb P$ represents a transition from the state at its bottom to the state at its top. In the fermionic $\chi$-directions, this transition is generated by a super-conformal symmetry of the $(j-1)$-th, $j$-th and $(j+1)$-th edges in this figure.}\label{squarepentagontransition}
\end{figure}

We now turn to the construction of the map. The question we should ask ourselves is: 
{\it What does it mean to charge a pentagon transition?} Said differently, how do we move from one pentagon-component to another in the $\chi$ decomposition of $\mathbb{P}$ in~(\ref{superP})? To find out, it helps thinking of the $\chi$'s as fermionic coordinates of sort and recall how usual (meaning bosonic) variables are dealt within the OPE set up.

The bosonic cross ratios are naturally associated with the symmetries of the middle squares. Namely, we can think of any middle square as describing a transition between two flux-tube states, one at its bottom and the other one at its top, as depicted in~\ref{squarepentagontransition}.\textbf{a}. Attached to this square are three conformal symmetries that preserve its two sides (left and right). To move in the space of corresponding cross ratios $(\tau, \sigma, \phi)$ we act on the bottom state with these symmetries
\beq
\psi_{\textrm{bottom}} \quad \longrightarrow \quad e^{-H\tau +iP \sigma +iJ\phi}\,\psi_{\textrm{bottom}}\, .
\eeq
Equivalently, we could act with the inverse transformations on the state at the top ($\psi_\text{top}$), since these are symmetries of the left and right sides sourcing the flux. In other words, the OPE family of Wilson loops is obtained by acting on all the twistors below each middle square with the conformal symmetries of that square. 

Similarly, to move in the space of `fermionic coordinates' we should act with a supercharge. In contrast to the previous case, these are now associated to the pentagons in the OPE decomposition. A pentagon transition represents the transition between two flux-tube states, one on the bottom square and the other on the top square -- the transition being induced from the shape of the pentagon. So what we should do is to find the supercharge that preserves the three sides of the pentagon sourcing the two fluxes, i.e.,~the  sides $j-1$, $j$ and $j+1$ in figure \ref{squarepentagontransition}.\textbf{b}, and act with it on the state at its bottom ($\psi_\text{bottom}$) or, equivalently, with the inverse symmetry on the state at its top ($\psi_\text{top}$). There is precisely one chiral supercharge that does the job, as we now describe.

Recall that we have 16 chiral supercharges at our disposal, that is, $\cQ^a_A$  where~$A$ is an $R$-charge index and~$a$ is an~$SL(4)$ twistor index. By construction they annihilate the super loop $\cW$ on which they act as~\cite{Korchemsky:2010ut}
\beq\la{Qact}
\cQ^a_A=\sum_{i=-1}^{n-2} Z^a_i{\d\over\d\eta_i^A}\qquad\text{with}\qquad\cQ^a_A\,\cW=0\, .
\eeq
By definition, for a given supercharge not to act on, say, the $i$-th side of the super loop, we need the coefficient of $\d/\d\eta_i^A$ to vanish. This can be achieved by contracting the $SL(4)$ index~$a$ with a co-twistor $Y$ such that $Y\cdot Z_i=0$. In our case, since we want $\cQ$ to be a symmetry of the three sides of a pentagon, the co-twistor should be orthogonal to $Z_{j-1}$, $Z_j$ and $Z_{j+1}$. There is exactly one such co-twistor:
\beq\la{cotwistor}
Y_j\equiv Z_{j-1}\wedge Z_j\wedge Z_{j+1} \,. 
\eeq

It is then straightforward to define the operator $\d/\d{\chi_{j}^A}$ that charges the $j$-th pentagon. It acts as $Y_j\cdot \cQ^A$ on the state $\psi_\text{bottom}$ entering the $j$-th pentagon from the bottom or, equivalently, on what have created this state. In other words, $\d/\d{\chi_{j}^A}$ is defined as $Y_j\cdot\cQ^A$ in~(\ref{Qact}) but with the summation restricted to edges lying below the $j$-th pentagon:
\beq
\boxed{{\d\over\d\chi_{j}^A}\equiv {1\over({\bf j-1})_j\,({\bf j})_j\,({\bf j+1})_j}\sum_{i = -1}^{j-2}Y_j\cdot Z_i\,{\d\over\d\eta_i^A}}\, . \label{noName}
\eeq
Alternatively we could act on the state $\psi_\text{top}$ at the top of the pentagon by restricting the summation to edges lying above the $j$-th pentagon and flipping the overall sign. These two prescriptions yield the same result since the two actions differ by $Y_j\cdot \cQ^A$ where $\cQ^A$ is the full supercharge annihilating the super loop. 

The normalization factor multiplying the sum in~(\ref{noName}) needs some explanation. It is introduced to make $\d/\d{\chi_{j}^A}$ weight free. In other words, it is defined such as to remove the weight of the co-twistor $Y_{j}$ used to define our supercharge.  In our notation, $({\bf i})_j$ extracts the weight of the twistor $Z_i$ in the $j$-th pentagon. This operation is unambiguous once we require it to be \textit{local} with respect to the $j$-th pentagon, meaning that it should only make use of the five twistors of this pentagon. Indeed, given a pentagon $p$ with five twistors $Z_a,\dots,Z_e$, the unique conformally invariant combination carrying weight with respect to $a$ is given by 
\beq
{({{\bf a})_{p}}}^4=\frac{\<abcd\>\<cdea\>\<deab\>\<eabc\>}{\<bcde\>^3}  \,. \la{weightP0}
\eeq
Uniqueness is very simple to understand. If another such expression existed, its ratio with (\ref{weightP0}) would be a conformal cross-ratio, which of course does not exist for a pentagon. A nice equivalent way of thinking of the weight (\ref{weightP0}) is as the NMHV tree level amplitude for the corresponding pentagon, that is
\beq\la{mentagonNMHV}
({\bf i})_{j}^{-4} = W_\text{$j$-th \pentagon}^{(iiii)\text{ tree}}\, .
\eeq
(Stated like this, the idea of dividing out by such weights is not new, see discussion around (132) in \cite{data}.) Multiplying three such weights to make the normalization factor in~(\ref{noName}), we would get 

\beq\la{totalweight}
\(({\bf j-1})_j\,({\bf j})_j\,({\bf j+1})_j\)^4=\frac{\left\langle Z_{j-1},Z_{j+1},Z_{t|j},Z_j\right\rangle {}^3 \left\langle
   Z_j,Z_{b|j},Z_{j-1},Z_{j+1}\right\rangle {}^3}{\left\langle
   Z_{j+1},Z_{t|j},Z_j,Z_{b|j}\right\rangle  \left\langle
   Z_{b|j},Z_{j-1},Z_{j+1},Z_{t|j}\right\rangle  \left\langle
   Z_{t|j},Z_j,Z_{b|j},Z_{j-1}\right\rangle }\, ,
 \eeq
where $Z_{t|j}$/$Z_{b|j}$ refer to the top/bottom twistors of the $j$-th pentagon respectively. Equivalently, $Z_{t|j}$/$Z_{b|j}$ are the middle twistors of the $(j\!+\!1)$-th/$(j\!-\!1)$-th pentagons, see figure~\ref{pentagontransition}. For further discussion of these weights and their rewriting see appendix \ref{weightSec}.

Finally, there are two minor ambiguities in the above construction on which we should comment. One is the overall normalization of (\ref{weightP0}) or (\ref{mentagonNMHV}) which is not fixed by the symmetry argument above. The convention chosen here is equivalent to setting 
\beq
\<\cP_{1234}\>=\[{\<Z_0,Z_1,Z_2,Z_{-1}\>\over({\bf 0})_{1}({\bf 1})_1({\bf 2})_1}\]^4\cW_\text{pentagon NMHV}^{(-1,-1,-1,-1)}=1 \,.
\eeq
A second minor ambiguity comes from the fourth power in (\ref{weightP0}) or (\ref{mentagonNMHV}). Due to its presence, to extract any weight we need to compute a fourth root, giving rise to a ${\mathbb Z}_4$ ambiguity. In practice we start from a point where the right hand side of (\ref{totalweight}) is real and positive for any $j$ and pick the positive fourth root when extracting the weight on the left. Then everything is real and can be nicely matched against the integrability predictions. This seems reminiscent of the sort of positivity regions of \cite{ArkaniHamed}. It would be interesting to study the $\mathbb{Z}_4$ ambiguity further, and possibly establish a connection to the positivity constraints of \cite{ArkaniHamed}.

\subsection{Interlude : Sanity Check}
\begin{figure}[t]
\centering
\def\svgwidth{3.5cm}
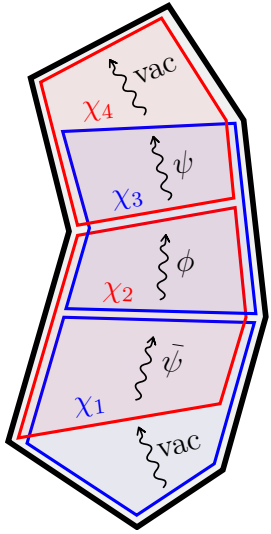
\caption{Leading OPE contribution to the NMHV octagon component $\cP_1 \circ \cP_2 \circ \cP_3 \circ \cP_4 = \frac{\partial}{\partial \chi_1^1}  \frac{\partial}{\partial \chi_2^2} \frac{\partial}{\partial \chi_3^3} \frac{\partial}{\partial \chi_4^4} \mathcal{W}$. For this component, each of the four pentagons in the octagon decomposition carries one unit of $R$-charge and fermion number. From the flux tube point of view, this corresponds to the sequence of transitions in equation (\ref{sequance}).}
\label{octagon1234}
\end{figure}

As a check of our map (\ref{noName}) we consider an eight-leg scattering amplitude, i.e., an octagon or, equivalently, a sequence of four pentagons. For concreteness, we focus on the example of~$\cP_1 \circ \cP_2 \circ \cP_3 \circ \cP_4= \frac{\partial}{\partial \chi_1^1}  \frac{\partial}{\partial \chi_2^2} \frac{\partial}{\partial \chi_3^3} \frac{\partial}{\partial \chi_4^4} \mathcal{W}$ at tree level and evaluate it in terms of the nine OPE variables $\{\tau_i,\sigma_i,\phi_i\}$. At this order, the OPE ratio $\mathcal{W}$ coincides with the ratio function $\mathcal{R}$ and we can easily extract components of the latter from the package \cite{Bourjaily:2013mma}.
For large OPE times we find that 
\beq
\cP_1 \circ \cP_2 \circ \cP_3 \circ \cP_4  = e^{-\tau_1-i \phi_1/2}\times e^{-\tau_2}\times e^{-\tau_3+i \phi_3/2} \times f(\sigma_i)+ \dots \la{example}
\eeq
which is actually already a non-trivial check of our construction. 
Indeed, we have four charged pentagons each of which injects one unit of R-charge and one unit of fermion number. As such, the lightest states that will flow in the three middle squares are a fermion $\bar \psi_{1}$ (with helicity $-1/2$) in the first square, a scalar $\phi_{12}$ (with no helicity) in the second square and the conjugate fermion $\psi_{123} = \psi^4$ (with helicity $+1/2$) in the last middle square. In short, the leading process contributing to this amplitude should correspond to the sequence of transitions 
\beq\la{sequance}
\text{vacuum} \stackrel{\cP_1}{\longrightarrow} \bar \psi_{1} \stackrel{\cP_2}{\longrightarrow} \phi_{12} \stackrel{\cP_3}{\longrightarrow} \psi_{123} \stackrel{\cP_4}{\longrightarrow} \text{vacuum} \,,
\eeq
as represented in figure \ref{octagon1234}. The three exponential factors in (\ref{example}) are in perfect agreement with this expectation.

Most importantly, the function $f(\sigma_i)$ should be given by the multiple Fourier transform of the sequence of pentagon transitions. It beautifully is. This and other similar checks -- at tree level and at loop level -- will be the subject of a separate longer publication \cite{ToAppear} whose main goal will be to precisely confront the program advocated here against the available perturbative data for non-MHV amplitudes.

\subsection{The Inverse Map} \la{inverseMap}
It is rather straightforward to invert the map (\ref{noName}) such as to obtain the $\d/\d\eta$'s in terms of the $\d/\d\chi$'s. For that aim, it is convenient to put back the weights in (\ref{noName}) and define
\beq
\cD^{(j)}_A\equiv ({\bf j-1})_j\,({\bf j})_j\,({\bf j+1})_j\,{\d\over\d\chi_j^A}=Y_j\cdot\sum_{i =-1}^{j-2} Z_i\,{\d\over\d\eta_i^A} \,. \la{defD}
\eeq
Given the triangular nature of this map, charging the first few edges at the bottom is as easy as writting the first few $\cD$'s explicitly. For the bottom edge, for instance, we immediately find that
\beq\la{deta1}
\cD^{(1)}_{A}=Y_1\cdot Z_{-1}{\d\over\d\eta^{A}_{-1}} \qquad \Rightarrow \qquad {\d\over\d\eta^{A}_{-1}}={\cD^{(1)}_{A}\over Y_1\cdot Z_{-1}}\, ,
\eeq
while taking this into account and moving to the following edge yields
\beq\la{deta2}
{\d\over\d\eta^{A}_0}={(Y_1\cdot Z_{-1})\cD^{(2)}_{A}-(Y_2\cdot Z_{-1})\cD^{(1)}_{A}\over (Y_1\cdot Z_{-1})(Y_2\cdot Z_{0})}\, ,
\eeq
and so on.

By following this recursive procedure we will eventually find that $\d/\d\eta_j$ is given as a linear combination of $\cD^{(j+2)}\,,\ \cD^{(j+1)}\,,\ \dots\,,\ \cD^{(1)}$. In plain words, it means that charging the edge $j$ entails charging the entire sequence of pentagons lying all the way from that specific edge to the bottom of the polygon. The drawback is that it has be so even for an edge standing arbitrarily far away from the bottom of the polygon. This, however, is at odds with the locality of the OPE construction, in which a random pentagon in the decomposition only talks to its neighbours (through the flux-tube state that they share) and has little knowledge of how far it stands from the bottom. Besides, it introduces an artificial discrimination between bottom and top, despite the fact that our analysis could, at no cost, be run from the top. The way out is easy to find: the bottom tail of the inverse map is pure mathematical illusion, or, put differently, the inverse map beautifully truncates such as to become manifestly top/bottom symmetric.
 
\begin{figure}[t]
\centering
\def\svgwidth{7cm}
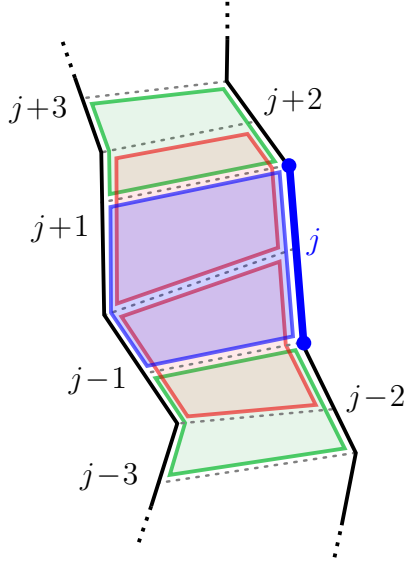
\caption{A remarkable {feature}
of our construction is that the inverse map turns out to be local. Namely, charging edge $j$ is done by charging the five pentagons touching this edge and these five pentagons alone.  We notice in particular that the two outermost pentagons in this neighbourhood, which are shown in green above, are touching the endpoints of edge $j$ only. 
}\label{figinversemap}
\end{figure}

In sum, instead of a sum over $j+2$ $\cD$'s, what we find is that (for $3\le j\le n-2$) $\d/\d\eta_j$ is given by the linear combination of the five neighboring pentagons only (see figure \ref{figinversemap})
\beq\la{inversemap}
\boxed{{\d\over\d\eta_j^A}={\<Y_{j-2},Y_{j-1},Y_{j},Y_{j+1}\>\,\cD^{(j+2)}_A+\quad\dots\quad+\<Y_{j-1},Y_{j},Y_{j+1},Y_{j+2}\>\,\cD^{(j-2)}_A\over \(Y_{j-1}\cdot Z_{j+1}\)\(Y_{j+1}\cdot Z_{j-1}\)\(Y_{j-2}\cdot Z_j\)\(Y_{j+2}\cdot Z_j\)}}\, .
\eeq

Mathematically, this relation originates from the five-term identity
\beq
\<Y_{j-2},Y_{j-1},Y_{j},Y_{j+1}\>Y_{j+2}+\quad\dots\quad+\<Y_{j-1},Y_{j},Y_{j+1},Y_{j+2}\>Y_{j-2}=0\, ,
\eeq
which holds for any choice of five (co-)twistors and which simply follows from them having four components. Once we plug the definition (\ref{defD}) into the right hand side of (\ref{inversemap}), most terms cancel out because of this identity. Those that survive are boundary terms and it is straightforward to work them out in detail. They precisely lead to the single term in the left hand side of (\ref{inversemap}).\footnote{To see that only the term proportional to $\partial/\partial \eta_j$ survives it is useful to note that the orthogonality relations $Y_{j-2}\cdot Z_{j-1}=Y_{j-1}\cdot Z_{j-1}=Y_{j}\cdot Z_{j-1}=0$ allow us to freely extend slightly the summation range of some of the five terms in (\ref{inversemap}). In turn, these relations follow trivially from the definition (\ref{cotwistor}) of the co-twistors. Finally, to check the overall normalization of both sides in (\ref{inversemap}), it is convenient to use the identity $\<Y_{j-2},Y_{j-1},Y_{j},Y_{j+1}\>= \(Y_{j-1}\cdot Z_{j+1}\)\(Y_{j+1}\cdot Z_{j-1}\)\(Y_{j-2}\cdot Z_j\)$.}

Actually, it is possible to interpret the inverse map (\ref{inversemap}) such that it also applies to the very first edges of the polygon, like in~(\ref{deta1}) and~(\ref{deta2}), provided that we properly understand what we mean by $Y_0$, $Y_{-1}$, $Y_{-2}$ and $Y_{-3}$. (These co-twistors will show up when using (\ref{inversemap}) for $\d/\d\eta_{2}$, $\d/\d\eta_{1}$, $\d/\d\eta_{0}$ and $\d/\d\eta_{-1}$.) For this we can pretend that there are extra edges at the bottom of the polygon and the previous derivation would still go through.\footnote{We can simply define $\(Y_0,Y_{-1},Y_{-2},Y_{-3}\)\equiv\(Y_{\{0,-1,1\}},Y_{\{*,0,-1\}},Y_{\{-1,*,*\}},Y_{\{*,*,*\}}\)$, with $Y_{\{i,j,k\}}\equiv Z_i\wedge Z_j\wedge Z_k$ and $Z_*$ being arbitrary twistors, which drop out of the final result. At the same time, we also set  $\(\cD_0\cW,\,\cD_{-1}\cW,\,\cD_{-2}\cW,\,\cD_{-3}\cW\)=\(0,0,0,0\).$}
Of course, for these bottom (or top) cases, it is easier to proceed recursively as in (\ref{deta1}) and (\ref{deta2}). 

This concludes our general discussion of the map. The proposals (\ref{inversemap}) and (\ref{noName}) are the main results of this paper.

\subsection{Easy Components and the Hexagon} \la{easySec}

A polygon with $n$ edges has a top pentagon and a bottom pentagon, plus $n-6$ pentagons which are neither top nor bottom and referred to as middle ones. Charging the bottom or the top pentagons is considerably simpler than charging any middle one. Let us focus on the bottom since the top is treated analogously. According to our general map (\ref{noName}), we see that the differential operator that charges the bottom pentagon, $\d/\d\chi_1$, is simply proportional to~$\partial/\partial \eta_{-1}$, 
\beq
{\d\over\d\chi_1}=\frac{Y_1\cdot Z_{-1}}{({\bf 0})_1 ({\bf 1})_1 ({\bf 2})_1} \times \frac{\partial}{\partial\eta_{-1}}
\eeq 
which we can further simplify to (see e.g. (\ref{toEvaluate}) in the appendix for a thorough explanation)
\beq
{\d\over\d\chi_1}=({\bf -1})_1 \times \frac{\partial}{\partial\eta_{-1}} \,.
\eeq 
In other words, up to a trivial factor which absorbs the weight in $\d/\d\eta_{-1}$, charging a bottom pentagon is the same as extracting components with $\eta$'s at the very bottom of our polygon. Similarly, charging the top-most pentagon is equivalent to putting $\eta$'s on the topmost edge. 
It could hardly be simpler. Explicitly, for any polygon, there are five NMHV components which are easy to construct:
\beqa
\cP_{1234}\circ\cP\circ \dots \circ \cP \circ \cP&=&w[4]\, \cW^{(-1,-1,-1,-1)} \nn\,,\\
\cP_{123}\circ\cP\circ \dots \circ \cP \circ \cP_{4}&=&w[3]\, \cW^{(-1,-1,-1,n-2)} \nn\,,\\
\cP_{12}\circ\cP\circ \dots \circ \cP \circ \cP_{34} &=&w[2]\, \cW^{(-1,-1,n-2,n-2)} \,, \la{easy} \\
\cP_{1}\circ\cP\circ \dots \circ \cP \circ \cP_{234}&=&w[1]\, \cW^{(-1,n-2,n-2,n-2)} \nn \,,\\
\cP_{}\circ\cP\circ \dots \circ \cP \circ \cP_{1234} &=&w[0]\, \cW^{(n-2,n-2,n-2,n-2)} \nn \,.
\eeqa
where $w[m]\equiv (({\bf -1})_1)^m\(({\bf n-2})_{n-4}\)^{4-m}$.

\begin{figure}[t]
\centering
\def\svgwidth{16cm}
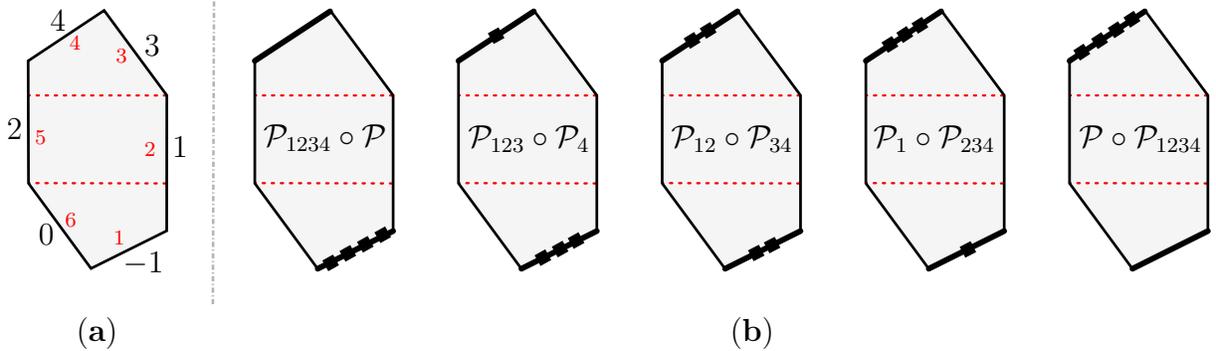
\caption{$(\bf{a})$ OPE friendly edge labelling used in this paper (big black outer numbers) versus the more conventional cyclic labelling (small red inner numbers)  for the hexagon. $(\bf{b})$ The five \textit{easy components} of the NMHV hexagon. Each black square represents a dual Grassmann variable $\eta$. For the hexagon these five components provide a complete base for all NMHV amplitudes.}\label{hexagons}
\end{figure}
These are what we call the \textit{easy components}. Morally speaking, from the first to the last line, we can think of the easy components as inserting an $F$, $\psi$, $\phi$, $\bar \psi$, $\bar F$ excitation and their conjugate at the very bottom and top of our polygon. 

For an hexagon we have only two pentagons and thus the easy components in (\ref{easy}) with $n=6$ suffice to describe the NMHV hexagon, see figure~\ref{hexagons}. All other components can be trivially obtained by Ward identities. For example, we can use invariance under
\beq
Y_2\cdot\cQ= \sum_k Y_2\cdot Z_k \frac{\partial}{\partial\eta_k} = 
 Y_2\cdot Z_{-1} \frac{\partial}{\partial\eta_{-1}}+ Y_2\cdot Z_{0} \frac{\partial}{\partial\eta_0}+ Y_2\cdot Z_{4} \frac{\partial}{\partial\eta_4}
\eeq
to replace any component with an index associated to the edge $0$ to a linear combination of components with $\eta$'s associated to the top and bottom edges $-1$ and $4$. Those, in turn, are the components which we can neatly compute from the OPE construction. For example, it immediately follows that 
\beqa
\la{LeftRule} \cW^{(-1,-1,-1,0)}=\alpha\, \cP_{1234}\circ \cP+ \beta \, \cP_{123}\circ \cP_4 \,, 
\eeqa
with 
\beqa
\alpha=- \frac{ Y_2\cdot Z_{-1}}{ Y_2\cdot Z_{0} \(({\bf -1})_1\)^4} \,\,\, , \qquad \beta=- \frac{Y_2\cdot Z_{4}}{ Y_2\cdot Z_{0} \(({\bf -1})_1\)^3\({\bf 4}_2\)^1} \,.
\eeqa
Similarly, we can easily write down any other hexagon NMHV component in terms of the OPE basis. Of course, this is equivalent to using the general inverse map (\ref{inversemap}), worked out in the previous section.

There are other components whose OPE expansions closely resemble those of the \textit{nice} components~(\ref{easy}). A notable example is the so-called \textit{cusp-to-cusp} hexagon scalar component $\mathcal{R}_{-1,0,3,4}^\text{hex}$ and its heptagon counterpart $\mathcal{R}_{-1,0,4,5}^\text{hep}$. Such components were extensively analyzed in the past using the OPE \cite{superOPE,withT2,data,andrei1}.\footnote{Recall once again that we are using here a slightly unconventional labelling of the edges as indicated in figure \ref{pentagontransition}.b; These same cusp-to-cusp component were denoted $\mathcal{R}_{6134}$ and $\mathcal{R}_{7145}$ in \cite{data} and $\mathcal{R}_{2356}$ in \cite{superOPE}.} What is nicest  about them is their utter simplicity at tree level, being described by a simple scalar propagator from the bottom cusp $(-1,0)$ to the top cusp $(n-3,n-2)$. Based on the OPE intuition, one would therefore imagine that this component should not behave that differently from the one in the middle line in (\ref{easy}). Indeed one observes that the expansion of this component and of the cusp-to-cusp components are exactly the same to leading order at large $\tau$ and to any loop order. Both are described by a single scalar flux tube excitation. However, as soon as two-particle contributions kick in -- in the sub-leading collinear terms -- these components start differing.  A similar story is present for all other components in the family~(\ref{easy}). For example, gluonic components were intensively studied in \cite{GluonPaper}. A simple tree-level gluonic example of this universality for the leading terms in the OPE was considered in detail in \cite{matheschool}. The hexagon fermonic component $\cW^{(-1,-1,-1,\,4)}$  was recently studied in \cite{andrei3}.\footnote{This component was denoted $\cW^{(1114)}$ in the cyclic labelling in \cite{andrei3}.} 

Finally let us note that the weight factors showing up in (\ref{easy}) are not a novelty. Already in \cite{data} it was explained that to properly deal with weight free quantities we better remove the weight of each pentagon by dividing out by the corresponding charged counterpart, see (132) and surrounding discussion in \cite{data}. Nevertheless, in practice, in all previous OPE studies of super amplitudes, the weights $({\bf -1})_1$ and $({\bf n-2})_{n-5}$ of the bottom and top twistors with respect to the corresponding pentagons were, for the most part, ignored. 
Sometimes this is fine. For instance, if we are interested in amplitudes at loop level we can always divide the ratio function by its tree level expression obtaining a weight free function of cross-ratios which we can unambiguously match with the OPE. Said differently, we can always normalize the tree level result by hand such that it agrees with the leading terms in the OPE. In particular, for the purpose of comparing with the hexagon function program \cite{Lance} and using the OPE to generate high loop order predictions, it is overkilling to carry these weights around. Moreover, with the choice of twistors in \cite{data}, such weights actually evaluate to $1$ which is one further reason why we never needed to take them into account. 

Having said all that, of course, to be mathematically rigorous, weight free quantities (\ref{easy}) are what we should always manipulate. In particular, for higher $n$-gons, and as soon as we also charge middle pentagons, it is important to keep track of these weights to properly make contact with the OPE predictions \cite{ToAppear}.

\subsection{Parity}

The charged pentagon construction provides us with a novel intuition about how to understand parity at the Wilson loop level. 

Recall that the action of parity on a scattering amplitude is very simple. It is a symmetry of the amplitude under which a positive helicity gluon transforms into a negative helicity one, a positive helicity fermion transforms into its negative helicity conjugate counterpart and finally a scalar excitation is trivially conjugated.  
All in all, this can be summarized in the following nice relation \cite{simplest} 
\beq
\int \prod_{i=-1}^{n-2} d^{4}  \tilde{\eta}_i \,\,e^{\,\sum_{i=-1}^{n-2}  \bar{\tilde{\eta}}_i \tilde{\eta}_i } A[\tilde\eta,\lambda,\tilde \lambda]=A[{\bar{\tilde\eta}},\tilde\lambda, \lambda]\,. \la{parityA}
\eeq
However, the relation between amplitudes and super Wilson loops involves stripping out the MHV tree-level factor along with going from the original amplitude $\tilde \eta$'s to the Wilson loop dual Grassman variables $\eta$'s.\footnote{The convention for the labelling of the $\eta$'s and $\tilde\eta$'s varies quite a lot in the literature. Our notation here is in line with \cite{superloopsimon} and \cite{ArkaniHamed} for example (modulo the non-cyclic labelling of the edges of course).} Together, these operations obscure the action of parity for this stripped object. How is parity symmetry realized on the super Wilson loop? Put differently, how does parity relate different ratio function components? To our knowledge this question has not been answered before. Here we propose that -- once decomposed using the OPE $\chi$-components
-- parity at the Wilson loop level is no more complicated than in the original amplitude language. Precisely, we claim that our variables allow for a straightforward analogue of (\ref{parityA}) in the Wilson loop picture as
\beq
\int \prod_{i=1}^{n-4} d^{4}  {\chi}_i \,e^{\,\sum_{i=1}^{n-4}  \bar{\chi}_i \chi_i } \mathcal{R}[\chi,Z]=\mathcal{R}[\bar{\chi},W] \,, \la{parityR}
\eeq
where $W_j$ are Hodge's \textit{dual} momentum twistors \cite{Hodges}. The latter can be thought of as parity conjugate of the $Z$'s and, up to an overall factor which drops out in (\ref{parityR}), are given by\footnote{
Dual momentum twistors (as well as usual momentum twistors) are reviewed in some detail in appendix \ref{variablesSec}, see formula (\ref{defW}) for an explicit expression relating them to the original momentum twistors, including all factors. Here we are dealing with weight free quantities and as a result, we can always safely drop any normalization from either $Z$'s or $W$'s on the left or right hand sides in (\ref{parityR}).}
\beq
W_j \equiv Z_{j-2} \wedge Z_j \wedge Z_{j+2} \,.
\eeq
Note that this is nothing but the conventional definition of the dual twistor involving three consecutive edges; the shifts of 2 in the index are just an outcome or our labelling, see figure~\ref{pentagontransition}. 

The general relation (\ref{parityR}) is a generating function for all parity relations between N$^k$NMHV components and N$^{n-4-k}$MHV components, such as the relation 
\beq
\cP_\Blue{1234}\circ\cP=\cP \circ\cP_\Red{1234}|_{Z\to W} \la{firstExample0}
\eeq
between two NMHV hexagon components, for instance, or the relation
\beq
\cP_\Blue{123}\circ\cP_\Red{14}\circ\cP_\Blue{234}=\cP_\Blue{4}\circ\cP_\Red{23}\circ\cP_\Blue{1}|_{Z\to W} \la{firstExample}
\eeq
relating NMHV and N$^2$MHV seven-point amplitudes. More precisely, to convert such identity into a relation for ratio function components, it is suffices to divide both sides by $\mathcal{W}=\cP\circ\cP\circ\cP$.\footnote{The latter is symmetric under $Z\to W$ so we can evaluate it with either twistors $Z$ or dual twistors $W$.} After doing so, the same relation in (\ref{firstExample}) reads
\beq
{\color{blue}\frac{\partial}{\partial \chi_1^{1}} \frac{\partial}{\partial \chi_1^{2}} \frac{\partial}{\partial \chi_1^{3}} }
{\color{red} \frac{\partial}{\partial \chi_2^{1}}  \frac{\partial}{\partial \chi_2^{4}}}
{\color{blue} \frac{\partial}{\partial \chi_3^{2}} \frac{\partial}{\partial \chi_3^{3}} \frac{\partial}{\partial \chi_3^{4}}} \,\mathcal{R}^\text{heptagon N$^2$MHV}
=\left.{\color{blue}  {\frac{\partial}{\partial \chi_1^{4}}} }
{\color{red}{ \frac{\partial}{\partial \chi_2^{2}} \frac{\partial}{\partial \chi_2^{3}}}}
{\color{blue} { \frac{\partial}{\partial \chi_3^{1}} }}\,\mathcal{R}^\text{heptagon NMHV} \right|_{Z\to W} \la{firstExampleBis}
\eeq
where the $\chi$ derivatives are given in terms of conventional $\eta$ derivatives in (\ref{noName}). Another more extreme example following from (\ref{parityR}) is the relation
\beq
\cP_\Blue{1234}\circ\cP_\Red{1234}\circ\cP_\Blue{1234}=\cP\circ\cP\circ\cP|_{Z\to W} \la{secondExample}
\eeq
which encodes the fact that for seven points N$^3$MHV is the same as $\overline{\text{MHV}}$. It is straightforward to generate more such relations by picking different components in (\ref{parityR}).   

Note that relations such (\ref{firstExampleBis}) are quite unconventional. We are not entitled to compare different $\eta$-components of the ratio function simply because they do not carry the same helicity weights. 
Equating different $\eta$-components would be tantamount to comparing apples and oranges. In contrast, when extracting the $\chi$-components as in (\ref{firstExampleBis}) we generate weight free quantities since the $\chi$'s -- contrary to the $\eta$'s -- carry no weight. This is what allows us to write parity relations at the level of the Wilson loop in terms of simple relations such as (\ref{firstExample})--(\ref{firstExampleBis}) or, simply, in terms of the master relation (\ref{parityR}), without the need of dressing the components by additional weight factors. 

Having decoded in detail the notation behind our main claim (\ref{parityR}), let us now explain how the relations  (\ref{firstExample})--(\ref{firstExampleBis}) are nicely suggested by the pentagon approach. Then, we will explain what sort of checks/derivations we have performed.

Parity, first and foremost, is a symmetry that swaps the helicity of the external particles in the $\mathcal{N}=4$ supermultiplet that are being scattered, see (\ref{parityA}). Similarly, parity also flips the helicity of the flux-tube excitations. 
Flipping the helicity of a flux-tube excitation is trivial: it can be accomplished by simply flipping the signs of all angles $\phi_j$'s, while keeping the times $\tau_j$ and distances $\sigma_j$ invariant \cite{moreOPE,short,data}. This is precisely what the transformation $Z \leftrightarrow W$ accomplishes!\footnote{More precisely, it is a very instructive exercise to observe that under $Z_j \rightarrow W_j$ the cross-ratios in formula (160) in \cite{data} precisely transform as $(\tau_j,\sigma_j,\phi_j) \to (\tau_j,\sigma_j,-\phi_j)$. When preforming such check it is important to take into account the conversion between the edge labelling used here and there, see caption of figure \ref{pentagontransition}.}

This explains the substitution rule in the right hand side of (\ref{firstExample0})--(\ref{secondExample}). To complete the picture we also have to act with parity on the pentagon transitions. Naturally, it is expected to swap the several super pentagon components in (\ref{superP}) in exactly the same way that it acts on the usual super-field multiplet expansion (replacing the positive helicity gluon with no $\tilde \eta$'s with the negative helicity gluon with 4 $\tilde \eta$'s and so on.). This translates into
\beq
\mathcal{P}_{1234} \leftrightarrow \mathcal{P} \,\,\, , \qquad \mathcal{P}_{123} \leftrightarrow \mathcal{P}_4 \,\,\, \text{etc},
\eeq
which is precisely what is encoded in (\ref{firstExample})--(\ref{firstExampleBis}) or, more generally, in (\ref{parityR}). In particular, these prescriptions neatly relate N$^k$MHV and N$^{n-k-4}$MHV amplitudes, as expected for parity. 

While (\ref{parityR}) is what the OPE naturally suggests, the previous paragraph is obviously not a proof. In any case, (\ref{parityR}) is a concrete conjecture for the realization of parity at the Wilson loop level that we should be able to establish (or disprove) rather straightforwardly starting from (\ref{parityA}), without any reference whatsoever to the OPE. It would be interesting if a simple and elegant derivation of (\ref{parityR}) existed, perhaps following the same sort of manipulations as in \cite{daveMason}. This would  elucidate further the origin of the (weight free) super OPE Grassmann variables $\chi$.

What we did was less thorough. To convince ourselves of the validity of (\ref{parityR}) we did two simpler exercises: On the one hand, using the very convenient package by Bourjaily, Caron-Huot and Trnka \cite{Bourjaily:2013mma} we extensively tested (\ref{parityR}) for a very large number of ratio functions from NMHV hexagons to N$^3$MHV decagons, both at tree and at one loop level.\footnote{When checking such identities for a very large number of edges, the package becomes unpractically slow. The trick is to open the package and do a ``find/replace operation'' to eliminate several \texttt{Simplify} and \texttt{FullSimplify} throughout. For analytical checks of relations such as (\ref{firstExample})--(\ref{firstExampleBis}), these simplifications are superfluous.} On the other hand, we also looked for an analytic derivation of (\ref{parityR}) from (\ref{parityA}). We did not find a particularly illuminating proof that establishes this in full generality but we did manage to prove several sub-examples. In appendix \ref{ParityAp}, for instance, we illustrate how one can rigorously establish the relation
\beq
\underbrace{\cP_{1234}\circ \dots \circ \cP_{1234}}_{j}\circ \underbrace{\cP_{} \circ \cP_{}}_{n-4-j} =\underbrace{\cP_{}\circ \dots \circ \cP_{}}_{j}\circ \underbrace{\cP_{1234} \circ \cP_{1234}}_{n-4-j}|_{Z \to W}\, .\label{parityP4...P}
\eeq

\section{Discussion}

In this paper we have constructed a simple map between N$^{k}$MHV amplitudes and so-called {\it charged pentagon transitions}. This map allows one to OPE expand amplitudes with arbitrary helicity configurations at any value of the coupling.  

In the dual super-loop description of the amplitude, the charged transitions are operators that act on the color flux tube. They can be realized as combinations of a properly chosen supercharge $\cal{Q}$ and the more standard bosonic pentagon operator $\cP$, and collected into a super pentagon

\beq\la{mathbbP}
{\mathbb P}=\cP\prod_{A=1}^4\(\One+\chi^AQ_A\)\, ,
\eeq
where the $\chi$'s are new (weight-free) Grassmann variables associated to the pentagons in the tessellation of the $n$-gon. The full super loop is then obtained by merging these pentagons together,
\beq\la{Psequence}
\cW_n=\<\mathbb{P}_1\circ\mathbb{P}_2\circ\dots\circ\mathbb{P}_{n-4}\>\, ,
\eeq
as previously done in (\ref{examples}). The $\chi$-components of the super transition ${\mathbb P}$ can be bootstrapped using the underlying flux-tube integrability, in pretty much the same way as their bosonic counterparts~\cite{data,andrei1,GluonPaper,andrei2}. In this paper we proposed a map between these $\chi$-components of the super Wilson loop $\cW_n$ and its more conventional $\eta$-components \cite{superloopskinner,superloopsimon}. This map, given in (\ref{noName}) and (\ref{inversemap}), therefore provides the key missing ingredient in the finite coupling OPE expansion of any helicity amplitude.

The map (\ref{noName}) can also be regarded as a definition of the charged transitions. In this construction, we charge a pentagon by acting with the corresponding super-symmetry generators on all the edges at its bottom (top). This is the same as acting on the flux-tube state entering the pentagon from the bottom (top) -- via the flux operator-state correspondence -- and is therefore equivalent to (\ref{mathbbP}).\footnote{This is pretty much the way the super loop was generated in \cite{superloopskinner,superloopsimon}.} 

This point of view is useful in providing a nice connection between the charged transitions and their non-charged counterparts. To illustrate this, consider a standard pentagon transition between a fermion and some other state, $P(\bar\psi_A(p)|\dots)=\<\dots|\cP|\bar\psi_A(p)\>$. As the momentum of this excitation, $p$, goes to zero, the fermion effectively becomes a super-symmetry generator $Q_A$ \cite{AldayMaldacena,BenDispPaper}. Therefore, we expect that in this limit this bosonic transition can be related to the charged transition $P_A(0|\dots)=\<\dots|\cP\cdot Q_A|0\>$. Indeed, while bootstrapping these transitions we have recently observed curious relations of the sort, 
\beq\la{psiat0p}
P_A(0|\dots) \propto  \oint\limits_{p=0} \frac{dp}{2\pi}\hat{\mu}_{\psi}(p)P(\bar{\psi}_A(p)|\dots)
\eeq
which seems to embody this idea in a rather sharp way.

We are currently exploring such directions and their generalizations and will present our findings elsewhere. Here we would like to briefly mention two interesting implications:

First, in the same way that we considered fermions $\bar \psi$ with zero momentum, we could also consider adding their conjugate $\psi$, which effectively becomes the conjugate super-symmetry generator $\bar \cQ$. This would naively define a non-chiral super pentagon admitting an expansion both in $\chi$'s and in $\bar \chi$'s. It is tempting to muse that it should be related to the non-chiral super loop proposed in \cite{simonHep} and further studied in \cite{Beisert:2012xx}.

Second, the relation~(\ref{psiat0p}) between $\cQ$ and $\bar\psi$ can be regarded as a local OPE definition of a charged edge or, equivalently, of the action of $\d/\d\eta$ on the super loop. 
Under such definition, our map (\ref{noName}) translates into a set of relations that includes the SUSY Ward identities, notably, and that begs to be interpreted directly from the flux tube. Naively, we expect them to encode certain discontinuities of the OPE series upon edge-crossing of the fermions. It would be fascinating to clarify this point and instructive to see if some simple OPE contour manipulation could provide a \textit{derivation} of supersymmetry from the flux tube theory.

Let us end with further outlook. There is by now a very large reservoir of knowledge on perturbative scattering amplitudes in planar $\mathcal{N}=4$ SYM theory, tightly related to the large amount of symmetries they are subject to (originating from both the original and dual Wilson loop descriptions). 
On the other hand, we have the pentagon approach, fully non-perturbative and valid all the way from weak to strong coupling.
This approach sacrifices some of the most basic symmetries of the amplitudes, such as supersymmetry, parity and cyclicity. In return, it renders the most non-trivial symmetry of all --  \textit{integrability} -- both manifest and practical. We think this is a worthy trade off, especially if the more conventional symmetries can be recovered in the end. Our map (\ref{noName}) is one realization of this philosophy, where different amplitudes that are related by supersymmetry are being assigned to the {\it same} OPE series. Moreover, as discussed above, we now start to understand that supersymmetry and parity also have, after all, a rather natural OPE incarnation. In our quest for the ultimate solution to the scattering amplitude problem, the next symmetry to attack is probably cyclicity.  Hopefully it will also turn out to be easier than we now think!

\section*{Acknowledgements}
We thank N.~Berkovits, J.~Bourjaily, S.~Caron-Huot, F.~Cachazo, J.~Maldacena and S.~He for numerous illuminating discussions, most notably concerning parity, and for inspiring remarks on the role of zero-momentum fermion for super amplitudes. We also thank the participants and organizers of the New Geometric Structures in Scattering Amplitudes program for an inspiring program and discussions. Research at the Perimeter Institute is supported in part by the Government of Canada through NSERC and by the Province of Ontario through MRI. The research of A.S. has been supported by the I-CORE Program of the Planning and Budgeting Committee, The Israel Science Foundation (grant No. 1937/12) and the EU-FP7 Marie Curie, CIG fellowship. L.C. and P.V. thank ICTP-SAIFR for warm hospitality during the concluding stages of this project. J.C. is funded by the FCT fellowship SFRH/BD/69084/2010. The research leading to these results has received funding from the People Programme (Marie Curie Actions) of the European Union's Seventh Framework Programme FP7/2007-2013/ under REA Grant Agreement No 317089 (GATIS). \textit{Centro de Fisica do Porto} is partially funded by the Foundation for 
Science and Technology of Portugal (FCT). 

\appendix
\section{More on Geometry, Pentagons and Parity}
In this appendix we review some known facts about the geometry of amplitudes and in particular, pentagons. These facts are then used in section \ref{ParityAp} to prove the parity relation~(\ref{parityP4...P}).  
\subsection{Variables} \la{variablesSec}
Scattering Amplitudes and null polygonal Wilson loops are conventionally parametrized by a plethora of very useful variables. Amongst them, we have momentum twistors $Z$, spinor helicity variables $\lambda$ and their parity conjugate $\tilde\lambda$, and dual momentum twistors $W$. Let us introduce them in our notation following \cite{Hodges} closely. We shall start by the momentum twistors $Z$ and construct all other variables from them. 

A momentum twistor is a four dimensional projective vector $Z_j \sim \lambda Z_j$. It is  associated to each edge of the null polygon, see figure \ref{pentagontransition}. Momentum twistors allow us to parametrize the shape of the polygon in an unconstrained way, this being one of their main virtues. Moreover, they transform linearly under conformal transformations and are therefore very useful when dealing with a conformal theory such as ${\cal N}=4$ SYM. 

Note the labelling of edges we are using in this paper is tailored from an OPE analysis and is \textit{not} the conventional cyclic labelling commonly used to describing color ordered partial amplitudes. In particular, in our convention, $Z_j$ and $Z_{j+ 1}$ (or $Z_{j-1}$) are \textit{not} neighbours; instead they nicely face each other in the polygon tessellation, see figure~\ref{pentagontransition}. The trivial conversion between our labelling and a more conventional numbering of the edges is presented in the caption of figure~\ref{pentagontransition}.

Out of four momentum twistors we can build conformal invariant angle brackets 
\beq
\<ijkl\>\equiv \epsilon_{abcd} Z_i^a Z_j^b Z_k^c Z_l^d \qquad \text{or} \qquad \<ijkl\> \equiv Z_i \wedge Z_j \wedge Z_k \wedge Z_l \, .
\eeq

We construct spinor helicity variables $\lambda$ by extracting the first two components of each four dimensional momentum twistors\footnote{
More precisely, we can always apply a global $GL(4)$ rotation $U$ to all the twistors (before extracting the first two components) plus a residual $GL(2)$ transformation $V$ to all the spinors (after extracting them from the first two components) such that in total $\lambda_{i}\equiv V\cdot \(\begin{array}{cccc} 1 & 0 & 0 & 0 \\ 0 & 1 & 0 & 0  \end{array} \)\cdot U\cdot Z_{i}$. Henceforth we set $U$ and $V$ to be the identity matrices. Nevertherless, it is worth keeping in mind that sometimes such transformations can be quite convenient. For instance, the twistors in previous OPE studies -- see e.g. appendix of \cite{data} -- contain several zero components and will lead to singular $\lambda$'s if extracted blindly. In those cases, it is quite convenient to preform such generic conformal transformations when constructing the spinor helicity variables.} 
\beq
\lambda_{i}\equiv \(\begin{array}{cccc} 1 & 0 & 0 & 0 \\ 0 & 1 & 0 & 0  \end{array} \)\cdot Z_{i}\, .
\eeq
With these spinor helicity variables we can construct Lorentz invariant two dimensional angle brackets
\beq
\<i,j\>\equiv \epsilon^{\alpha\beta} \lambda_{i,\alpha} \lambda_{j,\beta}  \qquad \text{or} \qquad \<i,j\>\equiv Z_i \cdot \mathbb{I}\cdot Z_{j}
\eeq
where $\mathbb{I}_{ab}$ is the usual infinite twistor which one can read off from the first definition. Next we introduce the dual momentum twistors $W$ which can be thought of as the parity conjugate of the $Z$'s. The dual momentum twistors are defined by using three neighbouring standard momentum twistors as
\beq
 W_{j,a}  \equiv \epsilon_{abcd} \frac{Z_{j-2}^b   Z_j^c   Z_{j+2}^d}{\<j-2,j\>\<j,j+2\>}
 \qquad \text{or} \qquad W_{j}  \equiv   \frac{Z_{j-2} \wedge Z_j  \wedge Z_{j+2}}{\<j-2,j\>\<j,j+2\>} \,. \la{defW}
\eeq
Note that with this convenient normalization the dual momentum twistor $W_j$ has the opposite helicity weight as the momentum twistor $Z_j$.
For the very bottom and top we need to tweak the definition (\ref{defW}) due to the non-cyclic labelling we are using.\footnote{Explicitly, the only tricky definitions are $W_{0}  \equiv \frac{Z_{2} \wedge Z_0 \wedge Z_{-1}}{\<2,0\>\<0,-1\>} $, $W_{-1}  \equiv \frac{Z_{0} \wedge Z_{-1} \wedge Z_{1}}{\<0,-1\>\<-1,1\>} $ at the bottom and $W_{n-2}  \equiv \frac{Z_{n-4} \wedge Z_{n-2} \wedge Z_{n-3}}{\<n-4,n-2\>\<n-2,n-3\>}$ and $W_{n-3}  \equiv \frac{Z_{n-2} \wedge Z_{n-3} \wedge Z_{n-5}}{\<n-2,n-3\>\<n-3,n-5\>}$  at the top, see figure~\ref{pentagontransition}.}

With the dual momentum twistors we can now construct four brackets once more, now denoted with square brackets
\beq
[ijkl]\equiv \epsilon^{abcd} W_{i,a} W_{j,b} W_{k,c} W_{l,d} \qquad \text{or} \qquad [ijkl] \equiv W_i \wedge W_j \wedge W_k \wedge W_l \,.
\eeq

Finally, we come to the parity conjugate spinor helicity variables $\tilde \lambda$.\footnote{Literally, the transformation $\(\lambda,\bar\lambda\)\to\(\bar\lambda,\lambda\)$ acts on the momentum $p_{\mu}\sigma^\mu_{\alpha \dot \alpha} = \lambda_\alpha \tilde \lambda_{\dot \alpha}$ as a reflection of $p_2$ since the corresponding Pauli matrix is antisymmetric while all others are symmetric. Once combined with an 180$^\circ$ rotation in the 1-3 plane, we get what is conventionally denoted by parity. In sum, since rotation symmetries are an obvious symmetry, one often slightly abuses notation to denote as parity any transformation whose determinant is $-1$. } 
They can be now defined as the \textit{last} two components of the dual twistors,
\beq
\tilde\lambda_{i}= \(\begin{array}{cccc} 0 & 0 & 1 & 0 \\ 0 & 0 & 0 & 1  \end{array} \)\cdot W_{i} \,.
\eeq
Out of two such twistors we can construct the Lorentz invariant square brackets
\beq
[ij] \equiv \epsilon^{\dot\alpha\dot\beta} \tilde\lambda_{i,\dot\alpha} \tilde\lambda_{j,\dot\beta}  \qquad \text{or} \qquad [ij]=W_i \cdot \tilde{\mathbb{I}} \cdot W_j
\eeq
where the dual infinity twistor $\tilde{\mathbb{I}}^{ab}$ can once gain be read off from the first definition. 

A beautiful outcome of the construction above is that momentum conservation
\beq
0= \sum_{i} \lambda_{i ,\alpha} \tilde\lambda_{i ,\dot\alpha}  \qquad \text{for $\alpha=1,2$ and $\dot \alpha=\dot 1,\dot 2$}
\eeq
automatically follows from the definitions above. In other words, as is well known, the use of twistors trivializes momentum conservation. 

To summarize: At this point, each edge of our polygon is endowed with a momentum twistor $Z_j$, a dual momentum twistor $W_j$ and a pair of spinors $\lambda_j$ and $\tilde \lambda_j$. There are also other null segments which play a critical role in our construction: the middle edges that define our tessellation which are represented by the red dashed lines in figure~\ref{pentagontransition} and whose corresponding momentum twistors are given in the caption of that same figure. We quote here for convenience: 
\beq
Z_\text{middle}=\<j-2,j,j+2,j-1\>Z_{j+1}-\<j-2,j,j+2,j+1\>Z_{j-1} \la{voila} \,.
\eeq
Let us briefly explain how this equation can be established. This simple exercise beautifully illustrates the power of Hodges' momentum twistors when dealing with the geometry of null lines. 
First, since $Z_\text{middle} \wedge Z_{j-1}$, $Z_\text{middle} \wedge Z_{j+1}$ and $Z_{j-1} \wedge Z_{j+1}$ all correspond to the same right cusp in figure~\ref{pentagontransition}a, we immediately have that $Z_\text{middle}=\alpha Z_{j+1}+\beta Z_{j-1}$. At the same time the point $Z_\text{middle} \wedge Z_{j}$ -- where the middle line intercepts the left edge in figure \ref{pentagontransition}a -- lies on the line $Z_{j+2}\wedge Z_{j}+t Z_{j-2}\wedge Z_{j}$ between the two left cusps. As such, the middle twistor is also a linear combination of the twistors $Z_{j}$, $Z_{j-2}$ and $Z_{j+2}$ and thus $\<j,j-2,j+2,Z_\text{middle}\>=0$. This condition immediately fixes the ratio $\beta/\alpha$ to be as in~(\ref{voila}). The normalization of the projective twistor can be fixed arbitrarily with~(\ref{voila}) being one such choice. Following the logic above, we can now also associate to each middle edge a dual twistor $W_\text{middle}$ and a pair of spinors $\lambda_\text{middle}$ and $\tilde \lambda_\text{middle}$. They will indeed show up below.

We close this section with two useful identities which we shall use latter. The first is 
\beq
\frac{\<i\hat{i}j\hat{j}\>}{[i\hat{i}j\hat{j}]}=\frac{\<i\hat{i}\>\<j\hat{j}\>}{[i\hat{i}][j\hat{j}]} \la{ratio1}
\eeq
where $\hat i$ and $i$ are neighbouring edges and so are $\hat j$ and $j$. The second is 
\beq
\<abcd\>=\<ab\>\<bc\>\<cd\>[bc] \qquad \text{and} \qquad [abcd]=[ab][bc][cd]\<bc\> \la{ratio2}
\eeq
which holds for any four consecutive twistors (starting with $a$ followed by $b$, then $c$ and then $d$ at the end). Note that the second equality in (\ref{ratio2}) follows from the first equality there together with (\ref{ratio1}). It also follows trivially from the first equality in (\ref{ratio2}) under parity which simply interchanges square and angle brackets.

\subsection{Pentagons and Weights} \la{weightSec}
In a tessellation of an $n$-sided polygon, each two consecutive null squares form a pentagon. As depicted in figure \ref{pentagontransition}, each such pentagon shares some edges with the larger polygon while some (either one or two) edges are middle edges defined by the tessellation, see also (\ref{voila}).

These pentagons play a prominent role in our construction. In particular, here we want to describe their importance in defining the \textit{weight} of a given edge \textit{with respect to a given pentagon}. To simplify our discussion we label the edges of a generic pentagon as $a,b,c,d,e$.\footnote{For example, this pentagon could be the first pentagon in the tessellation in figure \ref{pentagontransition}b. In this case we would set $a=Z_{2}$, $b=Z_0$, $c=Z_{-1}$, $d=Z_{1}$ and $e=Z_{\text{middle line ending on edge $2$}}$.} 

Pentagons have no cross-ratios. Nevertheless, they are not totally trivial. For instance, they allow us to read of the weight of an edge of the pentagon (with respect to that pentagon) through the pentagon NMHV ratio function components as
\beq
\mathcal{R}^{(abcd)}=\frac{1}{\bf a \bf b \bf c \bf d} \,\,\, , \qquad \mathcal{R}^{(aabc)}=\frac{1}{{\bf a}^2 \bf b \bf c} \,\,\,,\qquad  \mathcal{R}^{(aaaa)}=\frac{1}{{\bf a}^4} \,, \la{R5}
\eeq
and so on. All such components can be extracted from a single R-invariant beautifully written using momentum twistors in \cite{Hodges,daveMason},
\beq
\mathcal{R}^\text{NMHV pentagon}= \frac{\prod\limits_{A=1}^4 (\<abcd\> \eta_e^A+\<bcde\> \eta_a^A+\<cdea\> \eta_b^A+\<deab\> \eta_c^A+\<eabc\> \eta_d^A)}{\<abcd\>\<bcde\>\<cdea\>\<deab\>\<eabc\>}\,.
\eeq
From the relations (\ref{R5}) we read
\beq
{\bf a}^4=\frac{\<abcd\>\<cdea\>\<deab\>\<eabc\>}{\<bcde\>^3}  \,. \la{weightP}
\eeq
We can also re-write this relation using (\ref{ratio2}) as
\beq
{\bf a}^4 =\frac{\<ab\>^4\<ea\>^4}{\<ab\>\<bc\>\<de\>\<ea\> \<cd\>}  / \frac{[cd]^4}{[ab][bc][cd][de][ea]}\la{weightP 2brackets}
\eeq
where the familiar Parke-Taylor chains nicely show up. 
\begin{figure}[t]
\centering
\def\svgwidth{5cm}
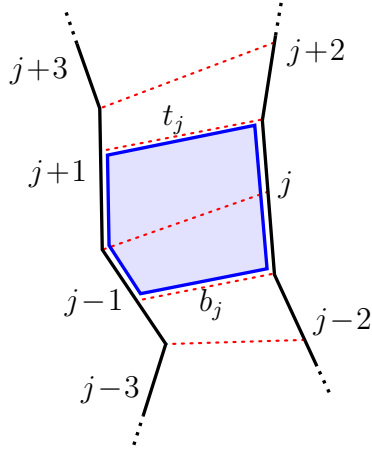
\caption{The weight factor ${\bf\color{black}(j-1)}_j({\bf\color{black}j})_j{\bf\color{black}(j+1)}_j$ associated to the $j$-th pentagon can be expressed in terms of the twistors of the larger polygon. It involves the seven closest edges to that pentagon, as illustrated in the figure. It is clear from this example the advantage of using this edge labelling as opposed to the cyclic one.}\label{weights}
\end{figure}

Furthermore, note that a product of three weights with respect to the same pentagon can be traded by the weight of any of the other two twistors of the pentagon using the first relation in (\ref{R5}) with $\mathcal{R}^{(abcd)} = 1/{\<abcd\>}$. In particular, it follows that
\beq
\frac{1}{\bf b \bf c \bf e}= \frac{ \bf a}{\<abce\>}= \frac{ \bf d}{\<dbce\>}\,.
\eeq
This allows us to massage slightly some of the formulae in the main text. For example, (\ref{noName}) can be written a bit more economically using
\beq
{1\over({\bf j-1})_j\,({\bf j})_j\,({\bf j+1})_j} = \frac{ ({\bf t_j})_j}{W_j \cdot Z_{t_j}}  \qquad \text{or}\qquad  {1\over({\bf j-1})_j\,({\bf j})_j\,({\bf j+1})_j} = \frac{ ({\bf b_j})_j}{W_j \cdot Z_{b_j}}  \la{toEvaluate}
\eeq
where $t_j$ and $b_j$ indicate the top or bottom twistors of pentagon $j$ respectively, see figure~\ref{weights}. Note that it is irrelevant that we do not fix the normalizations of these top and bottom twistors: they drop out in the ratios here constructed. 

We can also explicitly evaluate (\ref{toEvaluate}) by plugging in (\ref{weightP}) the expressions for the middle momentum twistors (\ref{voila}), see figure~\ref{weights}. When doing so, one finds\footnote{As usual, for the bottom and top pentagons we need to adjust this formula slightly. For instance, for $j=1$ we find $Z_{-2}$ in the right hand side which is not defined, see figure~\ref{pentagontransition}. The fix is very simple: we should simply replace $Z_{-2}$ by the very bottom twistor, that is $Z_{-1}$. 
Similarly for the top pentagon, where we should replace $Z_{n-1}$ by the very top twistor $Z_{n-2}$.}
\beqa\la{weightsexplicate}
&&\!\!\!\!\!\!\!\!\!\!\!\left(\!\frac{1}{{\bf\color{black}(j-1)}_j({\bf\color{black}j})_j{\bf\color{black}(j+1)}_j}\!\right)^4= \\ &&\,\,\,= \frac{\langle j-3,j-1,j+1,j+3\rangle  \langle j-2,j-1,j,j+2\rangle  \langle j-2,j,j+1,j+2\rangle }{\langle j-2,j-1,j,j+1\rangle ^2 \langle j-1,j,j+1,j+2\rangle ^2 \langle j-1,j,j+1,j+3\rangle  \langle j-3,j-1,j,j+1\rangle } \,.\nn
\eeqa

\subsection{Parity Map} \la{ParityAp}
In this section we establish the parity relation \eqref{parityP4...P} which we can equivalently cast as
\beq
\(\frac{\partial}{\partial{\chi_1}}\)^4 \dots \(\frac{\partial}{\partial{\chi_j}}\)^4 \mathcal{R}=\left.\(\frac{\partial}{\partial{\chi_{j+1}}}\)^4 \dots \(\frac{\partial}{\partial{\chi_{n-4}}}\)^4 \mathcal{R}\right|_{Z \to W} \, . \la{parityR2}
\eeq
To evaluate the left hand side we note that 
\beq
\frac{\partial}{\partial \chi_k}=\frac{\<k-1,k,k+1,k-2\>}{({\bf k-1})_k\,({\bf k})_j\,({\bf k+1})_k} \,\frac{\partial}{\partial\eta_{k-2}}+\dots \la{here2}
\eeq
where the $\dots$ contain a linear combination of derivatives from $\partial/\partial \eta_{-1}$ until $\partial/\partial \eta_{k-3}$. Since we are taking the maximum number of each derivative $\partial/\partial \chi_k$, only the term written in~(\ref{here2}) contributes, while all other terms are already saturated by the previous derivatives. Therefore, at the end we are left with a single ratio function component $$\mathcal{R}_{(-1)^4...\,(j-2)^4}\equiv \mathcal{R}^{(-1,-1,-1,-1),...,(j-2,j-2,j-2,j-2)}\,.$$ 
We can proceed in a similar fashion for the right hand side of (\ref{parityR2}), restricting the sum in \eqref{noName} to the edges above that pentagon. Keeping track of the multiplicative weight factors, we can now rewrite (\ref{parityR2}) as  
\beq
\frac{\mathcal{R}_{(-1)^4...\,(j-2)^4} }{\mathcal{R}_{(n-2)^4...\,(j+3)^4}\big |_{Z\to W}}= \( \prod\limits_{k={j+1}}^n \frac{\<k-1,k,k+1,k+2\>}{({\bf k-1})_k\,({\bf k})_k\,({\bf k+1})_k}\right)^4_{\!\!\!Z \to W} / \(\prod\limits_{k=1}^j \frac{\<k-1,k,k+1,k-2\>}{({\bf k-1})_k\,({\bf k})_k\,({\bf k+1})_k} \)^4 \, . \la{goal1}
\eeq

At this point, it is convenient to revert back to a more conventional cyclic notation. We shall revert to cyclic variables using the map in caption of figure \ref{pentagontransition} followed by a simple overall cyclic rotation of all the indices (by a convenient $n$ and $j$ dependent amount). Altogether, we map each edge index $k$ in (\ref{goal1}) as
\beq
k \to n - \frac{j}{2} - \frac{1}{2} \,\delta_{j\,\text{odd}}- \frac{k}{2}\, \delta_{k\,\text{even}}+\frac{k+3}{2}\, \delta_{k\,\text{odd}}\,.
\eeq
This change of labelling is illustrated in figure \ref{fig:parity octagon} for $n=8$ and $j=3$.
To avoid any confusions, we will add a $\mathcal{C}$ to the label of all equations written in this cyclic labelling.  Next, it is useful to convert the ratio in (\ref{goal1}) to two-brackets using \eqref{weightP 2brackets}, (\ref{ratio1}) and (\ref{ratio2}). In two-bracket notation, the parity transformation $Z\to W$ simply amounts to interchanging square and angle brackets. At the end of the day, we arrive at the nice expression \footnote{When simplifying the ratio of weights it is convenient to explore momentum conservation for the various middle squares to see that the dependence on the middle spinors neatly drops out. Recall that for any square momentum conservation $\sum_{j=1}^4 \lambda_j \tilde \lambda_j=0$ readily leads to $\<1,2\>[2,3]=-\<1,4\>[4,3]$ and $\<1,2\>[2,1]=\<3,4\>[3,4]$.}
\begin{Ceq}
\beq
\dfrac{\mathcal{R}_{(n)^4\ldots(n-j+1)^4}}{\mathcal{R}_{(n-j-2)^4\ldots(3)^4}|_{Z\to W}}=\dfrac{\<1,2\>\ldots\<n,1\>}{\<n,1\>^4\ldots\<n-j, n-j+1\>^4}\dfrac{[n-j-2, n-j-1]^4\ldots[2,3]^4}{[1,2]\ldots[n,1]} \, . \la{ratio parity} 
\eeq
To summarize: The main goal of this appendix is to establish this relation thus proving (\ref{parityP4...P}).  

\begin{figure}
\centering
\def\svgwidth{4cm}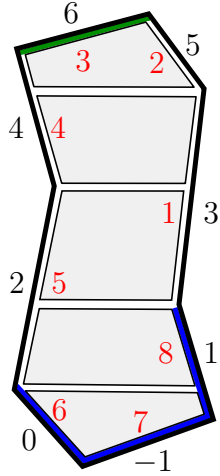
\caption{Example for N$^3$MHV octagon, ($n=8$, $j=3$). In blue, the edges charged for $\cP_{1234}\circ\cP_{1234}\circ\cP_{1234}\circ\cP$ and in green the edge charged for the parity conjugate $\cP\circ\cP\circ\cP\circ\cP_{1234}$. The OPE labelling for the edges is presented in black and in red the cyclic labelling used in this derivation.\label{fig:parity octagon}}
\end{figure}

To do so, we start with the amplitude picture where parity is very well understood -- it amounts to the exchange of $\lambda\leftrightarrow\tilde\lambda$ and the usual Grassmann variables $\tilde\eta\leftrightarrow\bar{\tilde\eta}$, see (\ref{parityA}). We shall show that (\ref{ratio parity}) is a simple consequence of the more transparent relation
\beq
\mathcal{A}_n[1^-,2^+,...\,,(n-j-1)^+,(n-j)^-,...\,,n^-]= \mathcal{A}_n[1^+,2^-,...\,,(n-j-1)^-,(n-j)^+,...\,,n^+]^*   \la{identityA} 
\eeq
for the scattering of $j+2$ negative helicity and $n-j-2$ positive helicity gluons. In a more supersymmetric notation, the quantity on the left hand side is the component $\(\tilde \eta_1\)^4 \(\tilde \eta_{n-j}\)^4 \dots \(\tilde \eta_{n}\)^4$ of the super amplitude
\beq
\mathcal{A}_n=\dfrac{\delta^8(\sum\limits_{i=1}^n \lambda_i\tilde\eta_i)}{\<1,2\>\ldots\<n,1\>}M_n^{\text{MHV loop}}(\lambda,\tilde\lambda) \mathcal{R}(\eta,Z) \, ,
\eeq
where $\mathcal{R}=1+\mathcal{R}_\text{NMHV}+\dots$ is the ratio function and $M_n^{\text{MHV loop}}(\lambda,\tilde\lambda)$ is the MHV amplitude divided by its tree level part. To extract this component we need to recall the relation between the Grassmann variables $\tilde \eta$ and $\eta$ showing up in this expression. In one direction, it reads \cite{daveMason}
\beq
\tilde\eta_i=\dfrac{\<i, i+1\>\eta_{i-1}+\<i+1, i-1\>\eta_{i}+\<i-1, i\>\eta_{i+1}}{\<i-1, i\>\<i, i+1\>}\la{parity amps} \,,
\eeq
while the inverse map is more subtle. It is not unique since we are working on the support of the supersymmetric delta function. In other terms, there is a gauge freedom which we can fix freely. One map that does the job gives $\eta_1=\eta_2=0$ and \cite{Bourjaily:2010wh,Korchemsky:2009hm}
\beqa
\eta_3&=&\<2,3\>\tilde\eta_2\, ,\nn\\
\eta_4&=&\<2,4\>\tilde\eta_2+\<3,4\>\tilde\eta_3\, ,\nn\\
&\vdots &\nn\\
\eta_n&=&\<2,n\>\tilde\eta_2+\ldots+\<n-1,n\>\tilde\eta_{n-1}\, .\la{tilde to etas}
\eeqa
Since $\tilde\eta_1$ and $\tilde\eta_n$ do not appear in this inverse map, we must look for them in the fermionic delta function when extracting this component. Therefore, we can simply replace the fermionic delta function by $ \<n,1\>^4$ and consider the component $ \(\tilde \eta_{n-j}\)^4 \dots \(\tilde \eta_{n-1}\)^4$ of the simpler quantity 
\beq
\left.\dfrac{\<n,1\>^4}{\<1,2\>\ldots\<n,1\>}M_n^{\text{MHV loop}}(\lambda,\tilde\lambda) \mathcal{R}(\eta,Z)\right|_{\eta_j=\sum_{k=2}^{j-1} \<k,j\> \tilde \eta_k} \, .
\eeq
In turn, this component is also straightforward to extract since it is another example of a saturation effect. More precisely, $\tilde \eta_{n-1}$ shows up only in $\eta_n$ such that extracting four units of it is tantamount to taking four powers of $\eta_n$ (times $\<n-1,n\>^4$). Next, $\eta_n$ is crossed out and $\tilde \eta_{n-2}$ shows up in $\eta_{n-1}$ only and so on. All in all, we arrive at
\beqa
&&\!\!\!\!\!\!\!\! \mathcal{A}_n[1^-,2^+,...\,,(n-j-1)^+,(n-j)^-,...\,,n^-]= \la{finL}\\&&\qquad \qquad\!\!\! =M_n^{\text{MHV loop}}(\lambda,\tilde\lambda)\, \dfrac{\<n,1\>^4}{\<1,2\>\ldots\<n,1\>}\<n-1, n\>^4\ldots\<n-j, n-j+1\>^4\,\mathcal{R}_{(n)^4\ldots\,(n-j+1)^4} \,.\nn
\eeqa
The right hand side of (\ref{identityA}) can be treated similarly.\footnote{When doing so it is convenient to note that the right hand side of (\ref{identityA}) can be also written as $(\mathcal{A}_n[1^-,2^+,...\,,(j+3)^+,(j+4)^-,...\,,n^-])^*_{i\rightarrow i-j-2}$. In this form, it is clear that we can simply recycle the result (\ref{finL}) with the obvious replacement of angle brackets by square brackets following from the conjugation.} In the end, we conclude that 
\beqa
&&\dfrac{\<n,1\>^4}{\<1,2\>\ldots\<n,1\>}\<n-1, n\>^4\ldots\<n-j, n-j+1\>^4\,\mathcal{R}_{(n)^4\ldots\,(n-j+1)^4}(Z)\nn\\
&=&\dfrac{[n-j-2,n-j-1]^4}{[1,2]\ldots[n,1]}[n-j-3, n-j-2]^4\ldots[2,3]^4\,\mathcal{R}_{(n-j-2)^4\ldots\,(3)^4}(W) \label{P4...P amp}
\eeqa
which gives precisely the ratio in \eqref{ratio parity} thus proving \eqref{parityP4...P}. This was the main goal of this appendix.  
\end{Ceq}

Other cases can be analyzed in a similar way. For instance, to establish an identity like $\cP_{1234}\circ\ldots\circ\cP_{1234}\circ\cP_{123}\circ\cP_{4}\circ\cP\circ\ldots\circ\cP=\cP\circ\ldots\circ\cP\circ\cP_{4}\circ\cP_{123}\circ\cP_{1234}\circ\ldots\circ\cP_{1234}|_{Z\to W}$ we start -- on the amplitude side -- with an amplitude that besides gluons also involves a positive helicity fermion $\psi$ and one negative helicity fermion $\bar \psi$. However, the analysis becomes more and more cumbersome as we consider cases with pentagons that are further away from being maximally charged. It would be interesting to streamline this analysis and work out the general case in a clean way. A better understanding of (the space of) all possible inverse maps $\eta(\tilde\eta)$ would probably be useful in this respect.

\end{document}

%% file: pentagontransition4.pdf_tex
\begingroup%
  \makeatletter%
  \providecommand\color[2][]{%
    \errmessage{(Inkscape) Color is used for the text in Inkscape, but the package 'color.sty' is not loaded}%
    \renewcommand\color[2][]{}%
  }%
  \providecommand\transparent[1]{%
    \errmessage{(Inkscape) Transparency is used (non-zero) for the text in Inkscape, but the package 'transparent.sty' is not loaded}%
    \renewcommand\transparent[1]{}%
  }%
  \providecommand\rotatebox[2]{#2}%
  \ifx\svgwidth\undefined%
    \setlength{\unitlength}{869.64257812bp}%
    \ifx\svgscale\undefined%
      \relax%
    \else%
      \setlength{\unitlength}{\unitlength * \real{\svgscale}}%
    \fi%
  \else%
    \setlength{\unitlength}{\svgwidth}%
  \fi%
  \global\let\svgwidth\undefined%
  \global\let\svgscale\undefined%
  \makeatother%
  \begin{picture}(1,0.44741877)%
    \put(0,0){\includegraphics[width=\unitlength]{pentagontransition4.pdf}}%
    \put(0.31103991,0.19595928){\color[rgb]{0,0,0}\makebox(0,0)[lb]{\smash{$\displaystyle{\psi}$}}}%
    \put(0.3111595,0.31480306){\color[rgb]{0,0,0}\makebox(0,0)[lb]{\smash{${\psi'}$}}}%
    \put(-0.00109522,0.25216215){\color[rgb]{0,0,0}\makebox(0,0)[lb]{\smash{$\displaystyle{\Blue{P(\psi|\psi')}}$}}}%
    \put(0.19896216,0.25275664){\color[rgb]{0,0,0}\makebox(0,0)[lb]{\smash{$j$}}}%
    \put(0.17713499,0.41390958){\color[rgb]{0,0,0}\makebox(0,0)[lb]{\smash{${j+2}$}}}%
    \put(0.16793581,0.0753797){\color[rgb]{0,0,0}\makebox(0,0)[lb]{\smash{${j-2}$}}}%
    \put(0.42609391,0.16717023){\color[rgb]{0,0,0}\makebox(0,0)[lb]{\smash{$ {j-1}$}}}%
    \put(0.4255129,0.35135515){\color[rgb]{0,0,0}\makebox(0,0)[lb]{\smash{$j+1$}}}%
    \put(0.2844867,0.26429745){\color[rgb]{0,0,0}\makebox(0,0)[lb]{\smash{$\Red{\text{middle}}$}}}%
    \put(0.94413944,0.12632505){\color[rgb]{0,0,0}\makebox(0,0)[lb]{\smash{$_1$}}}%
    \put(0.7951127,0.16128194){\color[rgb]{0,0,0}\makebox(0,0)[lb]{\smash{$_2$}}}%
    \put(0.93310042,0.19807867){\color[rgb]{0,0,0}\makebox(0,0)[lb]{\smash{$_3$}}}%
    \put(0.80431188,0.23119572){\color[rgb]{0,0,0}\makebox(0,0)[lb]{\smash{$_4$}}}%
    \put(0.8153509,0.07848931){\color[rgb]{0,0,0}\makebox(0,0)[lb]{\smash{$_0$}}}%
    \put(0.89814353,0.07112996){\color[rgb]{0,0,0}\makebox(0,0)[lb]{\smash{$_{-1}$}}}%
    \put(0.88894435,0.43909723){\color[rgb]{0,0,0}\makebox(0,0)[lb]{\smash{$_{n-2}$}}}%
    \put(0.78775336,0.40598017){\color[rgb]{0,0,0}\makebox(0,0)[lb]{\smash{$_{n-3}$}}}%
    \put(0.94965895,0.38022247){\color[rgb]{0,0,0}\makebox(0,0)[lb]{\smash{$_{n-4}$}}}%
    \put(0.76935499,0.32502738){\color[rgb]{0,0,0}\makebox(0,0)[lb]{\smash{$_{n-5}$}}}%
    \put(0.94597928,0.31398836){\color[rgb]{0,0,0}\makebox(0,0)[lb]{\smash{$_{n-6}$}}}%
    \put(0.27832599,0.00178625){\color[rgb]{0,0,0}\makebox(0,0)[lb]{\smash{$({\bf a})$}}}%
    \put(0.85235492,0.00178625){\color[rgb]{0,0,0}\makebox(0,0)[lb]{\smash{$({\bf b})$}}}%
  \end{picture}%
\endgroup%

%% file: calW3.pdf_tex
\begingroup%
  \makeatletter%
  \providecommand\color[2][]{%
    \errmessage{(Inkscape) Color is used for the text in Inkscape, but the package 'color.sty' is not loaded}%
    \renewcommand\color[2][]{}%
  }%
  \providecommand\transparent[1]{%
    \errmessage{(Inkscape) Transparency is used (non-zero) for the text in Inkscape, but the package 'transparent.sty' is not loaded}%
    \renewcommand\transparent[1]{}%
  }%
  \providecommand\rotatebox[2]{#2}%
  \ifx\svgwidth\undefined%
    \setlength{\unitlength}{985.85878906bp}%
    \ifx\svgscale\undefined%
      \relax%
    \else%
      \setlength{\unitlength}{\unitlength * \real{\svgscale}}%
    \fi%
  \else%
    \setlength{\unitlength}{\svgwidth}%
  \fi%
  \global\let\svgwidth\undefined%
  \global\let\svgscale\undefined%
  \makeatother%
  \begin{picture}(1,0.40012911)%
    \put(0,0){\includegraphics[width=\unitlength]{calW3.pdf}}%
    \put(-0.00269436,0.1895545){\color[rgb]{0,0,0}\makebox(0,0)[lb]{\smash{${\cal W}\equiv$}}}%
  \end{picture}%
\endgroup%

%% file: squarepentagontransition.pdf_tex
\begingroup%
  \makeatletter%
  \providecommand\color[2][]{%
    \errmessage{(Inkscape) Color is used for the text in Inkscape, but the package 'color.sty' is not loaded}%
    \renewcommand\color[2][]{}%
  }%
  \providecommand\transparent[1]{%
    \errmessage{(Inkscape) Transparency is used (non-zero) for the text in Inkscape, but the package 'transparent.sty' is not loaded}%
    \renewcommand\transparent[1]{}%
  }%
  \providecommand\rotatebox[2]{#2}%
  \ifx\svgwidth\undefined%
    \setlength{\unitlength}{436.42880859bp}%
    \ifx\svgscale\undefined%
      \relax%
    \else%
      \setlength{\unitlength}{\unitlength * \real{\svgscale}}%
    \fi%
  \else%
    \setlength{\unitlength}{\svgwidth}%
  \fi%
  \global\let\svgwidth\undefined%
  \global\let\svgscale\undefined%
  \makeatother%
  \begin{picture}(1,0.44938976)%
    \put(0,0){\includegraphics[width=\unitlength]{squarepentagontransition.pdf}}%
    \put(0.18245546,0.06399498){\color[rgb]{0,0,0}\makebox(0,0)[lb]{\smash{$\psi_\text{bottom}$}}}%
    \put(0.18245546,0.44160525){\color[rgb]{0,0,0}\makebox(0,0)[lb]{\smash{$\psi_\text{top}$}}}%
    \put(-0.00085049,0.2582993){\color[rgb]{0,0,0}\makebox(0,0)[lb]{\smash{$\text{left}$}}}%
    \put(0.36942754,0.2582993){\color[rgb]{0,0,0}\makebox(0,0)[lb]{\smash{$\text{right}$}}}%
    \put(0.75437005,0.06399498){\color[rgb]{0,0,0}\makebox(0,0)[lb]{\smash{$\psi_\text{bottom}$}}}%
    \put(0.73237333,0.44160525){\color[rgb]{0,0,0}\makebox(0,0)[lb]{\smash{$\psi_\text{top}$}}}%
    \put(0.61505752,0.25463318){\color[rgb]{0,0,0}\makebox(0,0)[lb]{\smash{$j$}}}%
    \put(0.94867436,0.1629802){\color[rgb]{0,0,0}\makebox(0,0)[lb]{\smash{$j-1$}}}%
    \put(0.94134212,0.34262003){\color[rgb]{0,0,0}\makebox(0,0)[lb]{\smash{$j+1$}}}%
    \put(0.20445218,0.00167096){\color[rgb]{0,0,0}\makebox(0,0)[lb]{\smash{$({\bf a})$}}}%
    \put(0.79103124,0.00167096){\color[rgb]{0,0,0}\makebox(0,0)[lb]{\smash{$({\bf b})$}}}%
  \end{picture}%
\endgroup%

%% file: octagon1234.pdf_tex
\begingroup%
  \makeatletter%
  \providecommand\color[2][]{%
    \errmessage{(Inkscape) Color is used for the text in Inkscape, but the package 'color.sty' is not loaded}%
    \renewcommand\color[2][]{}%
  }%
  \providecommand\transparent[1]{%
    \errmessage{(Inkscape) Transparency is used (non-zero) for the text in Inkscape, but the package 'transparent.sty' is not loaded}%
    \renewcommand\transparent[1]{}%
  }%
  \providecommand\rotatebox[2]{#2}%
  \ifx\svgwidth\undefined%
    \setlength{\unitlength}{186.89887695bp}%
    \ifx\svgscale\undefined%
      \relax%
    \else%
      \setlength{\unitlength}{\unitlength * \real{\svgscale}}%
    \fi%
  \else%
    \setlength{\unitlength}{\svgwidth}%
  \fi%
  \global\let\svgwidth\undefined%
  \global\let\svgscale\undefined%
  \makeatother%
  \begin{picture}(1,2.00358963)%
    \put(0,0){\includegraphics[width=\unitlength]{octagon1234.pdf}}%
    \put(0.58583846,0.27314836){\color[rgb]{0,0,0}\rotatebox{14.33913585}{\makebox(0,0)[lb]{\smash{$\text{vac}$}}}}%
    \put(0.58864245,0.60921614){\color[rgb]{0,0,0}\rotatebox{-8.41566933}{\makebox(0,0)[lb]{\smash{$\bar\psi$}}}}%
    \put(0.64886641,0.98299956){\color[rgb]{0,0,0}\makebox(0,0)[lb]{\smash{$\phi$}}}%
    \put(0.6418046,1.35366){\color[rgb]{0,0,0}\rotatebox{7.44596938}{\makebox(0,0)[lb]{\smash{$\psi$}}}}%
    \put(0.50205653,1.69121051){\color[rgb]{0,0,0}\rotatebox{21.17197008}{\makebox(0,0)[lb]{\smash{$\text{vac}$}}}}%
    \put(0.26397232,0.44832202){\color[rgb]{0,0,0}\rotatebox{7.18197892}{\makebox(0,0)[lb]{\smash{$\Blue{\chi_1}$}}}}%
    \put(0.40825355,1.23484074){\color[rgb]{0,0,0}\rotatebox{9.26668673}{\makebox(0,0)[lb]{\smash{$\Blue{\chi_3}$}}}}%
    \put(0.36873442,0.89116976){\color[rgb]{0,0,0}\rotatebox{-2.17291553}{\makebox(0,0)[lb]{\smash{$\Red{\chi_2}$}}}}%
    \put(0.29385362,1.57215504){\color[rgb]{0,0,0}\rotatebox{6.90589958}{\makebox(0,0)[lb]{\smash{$\Red{\chi_4}$}}}}%
  \end{picture}%
\endgroup%

%% file: inversemap.pdf_tex
\begingroup%
  \makeatletter%
  \providecommand\color[2][]{%
    \errmessage{(Inkscape) Color is used for the text in Inkscape, but the package 'color.sty' is not loaded}%
    \renewcommand\color[2][]{}%
  }%
  \providecommand\transparent[1]{%
    \errmessage{(Inkscape) Transparency is used (non-zero) for the text in Inkscape, but the package 'transparent.sty' is not loaded}%
    \renewcommand\transparent[1]{}%
  }%
  \providecommand\rotatebox[2]{#2}%
  \ifx\svgwidth\undefined%
    \setlength{\unitlength}{221.13088379bp}%
    \ifx\svgscale\undefined%
      \relax%
    \else%
      \setlength{\unitlength}{\unitlength * \real{\svgscale}}%
    \fi%
  \else%
    \setlength{\unitlength}{\svgwidth}%
  \fi%
  \global\let\svgwidth\undefined%
  \global\let\svgscale\undefined%
  \makeatother%
  \begin{picture}(1,1.06299119)%
    \put(0,0){\includegraphics[width=\unitlength]{inversemap.pdf}}%
    \put(0.56061278,0.59011926){\color[rgb]{0,0,0}\makebox(0,0)[lb]{\smash{$\color{blue}{j}$}}}%
    \put(0.10827982,0.32544585){\color[rgb]{0,0,0}\makebox(0,0)[lb]{\smash{$j\!-\!1$}}}%
    \put(0.03781719,0.61416381){\color[rgb]{0,0,0}\makebox(0,0)[lb]{\smash{$j\!+\!1$}}}%
    \put(0.48050079,0.85555546){\color[rgb]{0,0,0}\makebox(0,0)[lb]{\smash{$j\!+\!2$}}}%
    \put(-0.00406287,0.83991001){\color[rgb]{0,0,0}\makebox(0,0)[lb]{\smash{$j\!+\!3$}}}%
    \put(0.63704589,0.2915486){\color[rgb]{0,0,0}\makebox(0,0)[lb]{\smash{$j\!-\!2$}}}%
    \put(0.12977825,0.14201795){\color[rgb]{0,0,0}\makebox(0,0)[lb]{\smash{$j\!-\!3$}}}%
  \end{picture}%
\endgroup%

%% file: hexagons.pdf_tex
\begingroup%
  \makeatletter%
  \providecommand\color[2][]{%
    \errmessage{(Inkscape) Color is used for the text in Inkscape, but the package 'color.sty' is not loaded}%
    \renewcommand\color[2][]{}%
  }%
  \providecommand\transparent[1]{%
    \errmessage{(Inkscape) Transparency is used (non-zero) for the text in Inkscape, but the package 'transparent.sty' is not loaded}%
    \renewcommand\transparent[1]{}%
  }%
  \providecommand\rotatebox[2]{#2}%
  \ifx\svgwidth\undefined%
    \setlength{\unitlength}{3025.25449219bp}%
    \ifx\svgscale\undefined%
      \relax%
    \else%
      \setlength{\unitlength}{\unitlength * \real{\svgscale}}%
    \fi%
  \else%
    \setlength{\unitlength}{\svgwidth}%
  \fi%
  \global\let\svgwidth\undefined%
  \global\let\svgscale\undefined%
  \makeatother%
  \begin{picture}(1,0.28779429)%
    \put(0,0){\includegraphics[width=\unitlength]{hexagons.pdf}}%
    \put(0.08848179,0.08468871){\color[rgb]{0,0,0}\makebox(0,0)[lb]{\smash{$_\Red{1}$}}}%
    \put(0.11390028,0.15847053){\color[rgb]{0,0,0}\makebox(0,0)[lb]{\smash{$_\Red{2}$}}}%
    \put(0.09022026,0.23598669){\color[rgb]{0,0,0}\makebox(0,0)[lb]{\smash{$_\Red{3}$}}}%
    \put(0.05158296,0.24700923){\color[rgb]{0,0,0}\makebox(0,0)[lb]{\smash{$_\Red{4}$}}}%
    \put(0.02317921,0.16880618){\color[rgb]{0,0,0}\makebox(0,0)[lb]{\smash{$_\Red{5}$}}}%
    \put(0.04805184,0.1000379){\color[rgb]{0,0,0}\makebox(0,0)[lb]{\smash{$_\Red{6}$}}}%
    \put(0.09639937,0.05757315){\color[rgb]{0,0,0}\makebox(0,0)[lb]{\smash{$-1$}}}%
    \put(0.13713192,0.15376216){\color[rgb]{0,0,0}\makebox(0,0)[lb]{\smash{$1$}}}%
    \put(0.11391877,0.23826717){\color[rgb]{0,0,0}\makebox(0,0)[lb]{\smash{$3$}}}%
    \put(0.0260173,0.08098654){\color[rgb]{0,0,0}\makebox(0,0)[lb]{\smash{$0$}}}%
    \put(-0.00029698,0.16893431){\color[rgb]{0,0,0}\makebox(0,0)[lb]{\smash{$2$}}}%
    \put(0.0357007,0.25870162){\color[rgb]{0,0,0}\makebox(0,0)[lb]{\smash{$4$}}}%
    \put(0.21296986,0.16259497){\color[rgb]{0,0,0}\makebox(0,0)[lb]{\smash{$\mathcal{P}_{1234}\circ \mathcal{P}$}}}%
    \put(0.05745197,0.00058346){\color[rgb]{0,0,0}\makebox(0,0)[lb]{\smash{$(\bf{a})$}}}%
    \put(0.60063556,0.00058346){\color[rgb]{0,0,0}\makebox(0,0)[lb]{\smash{$(\bf{b})$}}}%
    \put(0.38301602,0.16212828){\color[rgb]{0,0,0}\makebox(0,0)[lb]{\smash{$\mathcal{P}_{123}\circ \mathcal{P}_4$}}}%
    \put(0.55180472,0.16222165){\color[rgb]{0,0,0}\makebox(0,0)[lb]{\smash{$\mathcal{P}_{12}\circ \mathcal{P}_{34}$}}}%
    \put(0.71988047,0.16233217){\color[rgb]{0,0,0}\makebox(0,0)[lb]{\smash{$\mathcal{P}_{1}\circ \mathcal{P}_{234}$}}}%
    \put(0.890731,0.16230916){\color[rgb]{0,0,0}\makebox(0,0)[lb]{\smash{$\mathcal{P}\circ \mathcal{P}_{1234}$}}}%
  \end{picture}%
\endgroup%

%% file: weights.pdf_tex
\begingroup%
  \makeatletter%
  \providecommand\color[2][]{%
    \errmessage{(Inkscape) Color is used for the text in Inkscape, but the package 'color.sty' is not loaded}%
    \renewcommand\color[2][]{}%
  }%
  \providecommand\transparent[1]{%
    \errmessage{(Inkscape) Transparency is used (non-zero) for the text in Inkscape, but the package 'transparent.sty' is not loaded}%
    \renewcommand\transparent[1]{}%
  }%
  \providecommand\rotatebox[2]{#2}%
  \ifx\svgwidth\undefined%
    \setlength{\unitlength}{181.77665506bp}%
    \ifx\svgscale\undefined%
      \relax%
    \else%
      \setlength{\unitlength}{\unitlength * \real{\svgscale}}%
    \fi%
  \else%
    \setlength{\unitlength}{\svgwidth}%
  \fi%
  \global\let\svgwidth\undefined%
  \global\let\svgscale\undefined%
  \makeatother%
  \begin{picture}(1,1.18523968)%
    \put(0,0){\includegraphics[width=\unitlength]{weights.pdf}}%
    \put(0.7112103,0.6737316){\color[rgb]{0,0,0}\makebox(0,0)[lb]{\smash{$j$}}}%
    \put(0.14334428,0.36055901){\color[rgb]{0,0,0}\makebox(0,0)[lb]{\smash{$j\!-\!1$}}}%
    \put(0.04882464,0.70298174){\color[rgb]{0,0,0}\makebox(0,0)[lb]{\smash{$j\!+\!1$}}}%
    \put(0.73564744,1.02420824){\color[rgb]{0,0,0}\makebox(0,0)[lb]{\smash{$j\!+\!2$}}}%
    \put(-0.00212235,0.97760148){\color[rgb]{0,0,0}\makebox(0,0)[lb]{\smash{$j\!+\!3$}}}%
    \put(0.804191,0.31052108){\color[rgb]{0,0,0}\makebox(0,0)[lb]{\smash{$j\!-\!2$}}}%
    \put(0.18710109,0.12861739){\color[rgb]{0,0,0}\makebox(0,0)[lb]{\smash{$j\!-\!3$}}}%
    \put(0.40290859,0.85299251){\color[rgb]{0,0,0}\makebox(0,0)[lb]{\smash{$t_j$}}}%
    \put(0.50049438,0.35743541){\color[rgb]{0,0,0}\makebox(0,0)[lb]{\smash{$b_j$}}}%
  \end{picture}%
\endgroup%

%% file: octagon.pdf_tex
\begingroup%
  \makeatletter%
  \providecommand\color[2][]{%
    \errmessage{(Inkscape) Color is used for the text in Inkscape, but the package 'color.sty' is not loaded}%
    \renewcommand\color[2][]{}%
  }%
  \providecommand\transparent[1]{%
    \errmessage{(Inkscape) Transparency is used (non-zero) for the text in Inkscape, but the package 'transparent.sty' is not loaded}%
    \renewcommand\transparent[1]{}%
  }%
  \providecommand\rotatebox[2]{#2}%
  \ifx\svgwidth\undefined%
    \setlength{\unitlength}{1008.71386719bp}%
    \ifx\svgscale\undefined%
      \relax%
    \else%
      \setlength{\unitlength}{\unitlength * \real{\svgscale}}%
    \fi%
  \else%
    \setlength{\unitlength}{\svgwidth}%
  \fi%
  \global\let\svgwidth\undefined%
  \global\let\svgscale\undefined%
  \makeatother%
  \begin{picture}(1,1.8)%
    \put(0,0){\includegraphics[width=\unitlength]{octagon.pdf}}%

    \put(0.5564173,0.10128257){\color[rgb]{0,0,0}\makebox(0,0)[lb]{\smash{$-1$}}}%
    \put(0.79036989,0.45817879){\color[rgb]{0,0,0}\makebox(0,0)[lb]{\smash{$1$}}}%
    \put(0.79036989,0.92451997){\color[rgb]{0,0,0}\makebox(0,0)[lb]{\smash{$3$}}}%
    \put(0.73036989,1.483451997){\color[rgb]{0,0,0}\makebox(0,0)[lb]{\smash{$5$}}}%
    \put(0.32583174,1.59179901){\color[rgb]{0,0,0}\makebox(0,0)[lb]{\smash{$6$}}}%
    \put(0.14246754,1.20666289){\color[rgb]{0,0,0}\makebox(0,0)[lb]{\smash{$4$}}}%
    \put(0.1453838,0.69227375){\color[rgb]{0,0,0}\makebox(0,0)[lb]{\smash{$2$}}}%
    \put(0.1864173,0.17128257){\color[rgb]{0,0,0}\makebox(0,0)[lb]{\smash{$0$}}}%

    \put(0.5564173,0.23128257){\color[rgb]{1,0,0}\makebox(0,0)[lb]{\smash{$7$}}}%
    \put(0.64036989,0.45817879){\color[rgb]{1,0,0}\makebox(0,0)[lb]{\smash{$8$}}}%
    \put(0.65036989,0.92451997){\color[rgb]{1,0,0}\makebox(0,0)[lb]{\smash{$1$}}}%
    \put(0.61036989,1.423451997){\color[rgb]{1,0,0}\makebox(0,0)[lb]{\smash{$2$}}}%
    \put(0.36583174,1.43179901){\color[rgb]{1,0,0}\makebox(0,0)[lb]{\smash{$3$}}}%
    \put(0.28246754,1.20666289){\color[rgb]{1,0,0}\makebox(0,0)[lb]{\smash{$4$}}}%
    \put(0.2853838,0.69227375){\color[rgb]{1,0,0}\makebox(0,0)[lb]{\smash{$5$}}}%
    \put(0.2864173,0.27128257){\color[rgb]{1,0,0}\makebox(0,0)[lb]{\smash{$6$}}}%

  \end{picture}%
\endgroup%